\documentclass[twocolumn]{aastex63}
\usepackage{appendix}
\usepackage{enumitem}
\usepackage{graphicx,url}
\usepackage{color}
\usepackage{amsmath}
\usepackage{multirow}
\usepackage[colorinlistoftodos]{todonotes}
\usepackage{xcolor}

\def\eso{ESO~362$\mbox{-}$G18}

\def\xmm{{\it XMM-Newton}}
\def\suzaku{{\it Suzaku}}
\def\chandra{{\it Chandra}}

\def\swift{{\it Swift}}
\def\nustar{{\it NuSTAR}}

\def\athena{{\it Athena}}
\def\eXTP{{\it eXTP}}
\def\hexp{{\it HEX-P}}
\def\gaussian{{\tt gaussian}}
\def\diskbb{{\tt diskbb}}
\def\tbabs{{\tt tbabs}}
\def\pl{{\tt powerlaw}}
\def\cutoffpl{{\tt cutoffpl}}
\def\xillver{{\tt xillver}}

\def\relxill{{\tt relxill}}

\def\relxillD{{\tt relxillD}}

\def\relxilllpD{{\tt relxilllpD}}
\def\relxillCp{{\tt relxillCp}}

\def\xstar{{\sc xstar}}

\def\nthComp{{\tt nthComp}}

\def\borus{{\tt borus12}}

\def\csq{$\chi^2$}

\def\afe{$A_{\mathrm{Fe}}$}
\def\logne{$\log{[n_\mathrm{e}/\mathrm{cm}^{-3}]}$}

\def\fluxunits{erg\,cm$^{-2}$\,s$^{-1}$}

\def\logxi{$\log{[\xi/\mathrm{erg}\,\mathrm{cm}\,\mathrm{s}^{-1}]}$}

\shorttitle{X-ray Reflection Spectroscopy of \eso}
\begin{document}

\title{\large\bf The Nature of Soft Excess in \eso\ Revealed by XMM-Newton and NuSTAR Spectroscopy}

\correspondingauthor{Yerong~Xu}
\email{yrxu047@gmail.com}

\author{Yerong~Xu}
\affil{Universit\`a degli Studi di Palermo, Dipartimento di Fisica e Chimica, via Archirafi 36, I-90123 Palermo, Italy}
\affil{INAF - IASF Palermo, Via U. La Malfa 153, I-90146 Palermo, Italy}

\author{Javier~A.~Garc\'ia}
\affil{Cahill Center for Astronomy and Astrophysics, California Institute of Technology, Pasadena, CA 91125, USA}
\affil{Dr. Karl Remeis-Observatory and Erlangen Centre for Astroparticle Physics, Sternwartstr.~7, 96049 Bamberg, Germany}

\author{Dominic~J.~Walton}
\affil{Institute of Astronomy, University of Cambridge, Madingley Road, Cambridge CB3 0HA, UK}

\author{Riley~M.~T.~Connors}
\affil{Cahill Center for Astronomy and Astrophysics, California Institute of Technology, Pasadena, CA 91125, USA}

\author{Kristin~Madsen}
\affil{Cahill Center for Astronomy and Astrophysics, California Institute of Technology, Pasadena, CA 91125, USA}

\author{Fiona~A.~Harrison}
\affil{Cahill Center for Astronomy and Astrophysics, California Institute of Technology, Pasadena, CA 91125, USA}

\begin{abstract}
We present a detailed spectral analysis of the joint \xmm\ and \nustar\ observations of the active galactic nuclei (AGN) in the Seyfert 1.5 Galaxy \eso. The broadband ($0.3\mbox{--}79$\,keV) spectrum shows the presence of a power-law continuum with a soft excess below $2$\,keV, iron K$\alpha$ emission ($\sim 6.4$\,keV), and a Compton hump (peaking at $\sim 20$\,keV). We find that the soft excess can be modeled by two different possible scenarios: a warm ($kT_\mathrm{e}\sim0.2\,\mathrm{keV}$) and optically thick ($\tau\sim34$) Comptonizing corona; or with relativistically-blurred reflection off a high-density (\logne$>18.3$) inner disk. 
These two models cannot be easily distinguished solely from their fit statistics. However, the low temperature ($kT_\mathrm{e}\sim20\,\mathrm{keV}$) and the thick optical depth ($\tau\sim5$) of the hot corona required by the warm corona scenario are uncommon for AGNs. We also fit a 'hybrid' model, which includes both disk reflection and a warm corona. Unsurprisingly, as this is the most complex of the models considered, this provides the best fit, and more reasonable coronal parameters. In this case, the majority of the soft excess flux arises in the warm corona component. However, based on recent simulations of warm coronae, it is not clear whether such a structure can really exist at the low accretion rates relevant for \eso\ ($\dot{m}\sim0.015$). This may therefore argue in favour of a scenario in which the soft excess is instead dominated by the relativistic reflection. Based on this model, we find that the data would require a compact hot corona ($h\sim3\,R_\mathrm{Horizon}$) around a highly spinning ($a_\star>0.927$) black hole.

\end{abstract}

\keywords{AGN: \eso, X-ray soft excess}

\section{Introduction} \label{sec:intro}

A wealth of observational evidence has indicated the existence of supermassive black holes (SMBHs) of mass $M_{BH} \sim 10^{6} \mbox{--} 10^{9.5}\,M_{\odot}$ located at the center of nearly all galaxies \citep[e.g., ][]{1995Kormendy,2013Kormendy}. Accretion of matter onto the SMBH is one of the most efficient mechanisms to transfer gravitational potential energy into electromagnetic radiation, covering a broad energy range from radio to X-rays, and even up to Gamma rays. Particularly, observations in the X-ray band can probe the very innermost regions of the accretion disk, as the bulk of the emitted flux originates in a centrally located and compact corona \citep[e.g.,][]{2015Fabian,2017Fabian}. Therefore, X-ray spectroscopy can reveal the properties of SMBHs and their interaction with the surroundings. The typical broad-band X-ray spectrum of a Seyfert AGN consists of a coronal power-law continuum, fluorescent emission features, a Compton hump and a soft excess below $\sim 2$\,keV. The power-law continuum often extends to high energies ending with a sharp cut-off, which is expected due to the inverse-Compton scattering of ultra-violet (UV) and optical photons from the accretion disk in the central hot corona \citep{1980Sunyaev,1993Haardt}. A fraction of the coronal continuum illuminates the accretion disk and gives rise to the Compton hump, fluorescent emission lines, and other reflection features \citep[e.g.,][]{2005Ross,2010Garc,2013Garc}. The Compton hump generally peaks at $20\mbox{--}30$\,keV, where the low-energy side is shaped by the photoelectric absorption of iron in the reflector, while the high-energy side stems from the Compton down-scattering of coronal high-energy photons reprocessed in the accretion disk or distant matter \citep[e.g.,][]{1990Pounds,1991Nandra}. The Fe K$\alpha$ line at $\sim 6.4$\,keV is the most notable atomic feature, which is usually broadened and distorted by relativistic effects due to the strong gravitational field around the black hole, if the reflection occurs in the inner-accretion disk \citep[e.g.,][]{1989Fabian,1991Laor}. The Fe K$\alpha$ line can also be observed to be narrower if the reflected emission originates farther from the black hole, for example, if produced in the broad line region or the distant torus \citep[e.g.,][]{1991George,1991Matt}. 

The soft excess is a broad and featureless spectral component commonly observed below $\sim 2$\,keV in the spectra of nearly half of the Seyfert AGNs \citep[e.g.,][]{1984Halpern, 1985Arnaud, 1989Turner}.
Its origin has been a controversial topic over the years. It was originally thought to be part of the UV blackbody emission from the AGN accretion disk \citep[e.g.,][]{1985Singh,1986Pounds,1999Leighly}. However, this explanation has been ruled out, because the corresponding disk temperature ($\sim 0.2$\,keV) is too high for a typical AGN accretion disk, which peaks in the UV band \citep[][hereafter SS73]{1973Shakura}. Moreover, AGNs with different accretion rates and masses are expected to have disks at very different temperatures, which is inconsistent with the narrow range of energies at which the soft excess is found in many sources \citep[e.g.,][]{2004Gierli,2005Piconcelli,2009Miniutti,2009Bianchi}. Currently, there are two competing  proposed models to explain the soft excess: a warm Comptonizing corona, and relativistically-blurred and ionized reflection. The first interpretation describes an scenario whereby the UV photons from the disk are Compton-scattered in a warm ($kT_{e} \sim 0.1\mbox{--}1$keV) and optically thick ($\tau \sim 10\mbox{--}40$) corona covering the inner regions of the disk, which is cooler than the hot central corona \citep[e.g.,][]{1987Czerny,2009Middleton,2012Done,2018Petrucci,2018Kubota,2019Ursini}. In the second model, the irradiated inner disk reprocesses the hard X-rays that originate from the hot corona, producing a multitude of fluorescent atomic lines, which are then blended and distorted by the relativistic blurring due to the strong gravity of the black hole \citep[e.g.,][]{2018Jiang,2019Garc}.

The Seyfert 1.5 galaxy \eso\ (a.k.a. MCG 05-13-17 or \swift\ J0501.9-3239) is a nearby AGN \citep[$z\sim0.012$;][]{2006Bennert}. This source has been observed by \xmm, \suzaku, and \chandra\ during $2005\mbox{--}2010$, displaying a clear short-timescale spectral variability \citep[][hereafter AG14]{2014Agis}. It was reported to host a rapidly spinning  black hole \citep[$a_{\star}>0.96$;][]{2013Walton},  with a mass of ($4.5\pm 1.5)\times10^{7}M_{\odot}$ (AG14). The bolometric luminosity was estimated at $L_\mathrm{bol}\sim1.3\times10^{44}~\mathrm{erg}\, \mathrm{s}^{-1}$, derived by AG14 by assuming the X-ray correction factor as $k_\mathrm{2-10}=25$, which translates to an Eddington-scaled accretion rate of $\dot{m}\equiv L_\mathrm{Bol}/L_\mathrm{Edd}\sim0.02$, where $L_\mathrm{Edd}=4\pi G M_\mathrm{BH}m_\mathrm{p}c/\sigma_{T}$. Estimations of the inclination angle between the line-of-sight and the axis of the broad-line region (BLR) and the galactic disk inclination are $i = 53^\circ\pm5^\circ$ and $i \approx37^\circ$ respectively \citep{2000Fraquelli,2018Humire}.  AG14 localised the X-ray emitting region to within $\sim50$ gravitational radii ($R_{\rm g}=GM/c^2$ ) through the absorption variability using $2005\mbox{--}2010$ multi-epoch X-ray observations. In this paper, we analyze \eso\ through the most recent \xmm\ \citep{2001Jansen} and \nustar\ \citep{2013Harrison} observations, investigating the presence of relativistic reflection features and the nature of the soft excess. The broad energy band ($\sim0.3\mbox{--}79$\,keV) provided by the combination of these two observatories enables us to study the X-ray reflection spectrum comprehensively.

The paper is organized as follows. In Section~\ref{sec:reduction}, we present the observational data reduction and the light curves. The spectral analysis with two different possible scenarios is reported in Section~\ref{sec:spectra}. We discuss the results and make conclusions in Section~\ref{sec:discussion} and Section~\ref{sec:conclusion} respectively.

\section{Data Reduction and Light curves} \label{sec:reduction}
The first \nustar\ observation of \eso\ was performed in September 2016 during \nustar\ Cycle 2 with an exposure time of $\sim102$\,ks, joint with a \xmm\ observation of $\sim121$\,ks. The detailed log of the observations is listed in Table~\ref{tab:obslog}. The fluxes in the $0.5\mbox{--}2$\,keV ($F_{0.5-2}$) and in the $2\mbox{--}10$\,keV ($F_{2-10}$) bands are also included. Compared with the archival data, where most of observations show $F_{0.5-2}>2\times10^{-12}\,\mathrm{erg\,\mathrm{s}^{-1}\,\mathrm{cm}^{-2}}$ and $F_{2-10}>10^{-11}\,\mathrm{erg\,\mathrm{s}^{-1}\,\mathrm{cm}^{-2}}$ (AG14), the present observations have a flux in the soft energy band that is much fainter, while that in the hard energy band remains at similar levels. Thus, unless absorption plays an important role, we expect a harder photon index and lower accretion rate than before.

%
\begin{deluxetable*}{lccccccc}[ht!]
\tablecaption{Observations log for \eso \label{tab:obslog}}
\tablecolumns{7}
\tablewidth{0pt}
\tablehead{
\colhead{Telescope} & \colhead{Instrument} & \colhead{Obs. ID} & \colhead{Date} &
\colhead{Net~Exp. (ks)} & \colhead{$F_{0.5-2}^{\star}$}  & \colhead{$F_{2-10}^{\star}$} }
\startdata
\nustar\  & FPMA/B   & 60201046002 & 2016-09-24 & 102 & \multirow{2}*{$0.85\pm0.01$} &\multirow{2}*{$10.5\pm0.1$}  \\
\xmm\    & EPIC  &  0790810101   & 2016-09-24 &  121 & ~ &~  \\
\enddata
\tablenotetext{\star}{The fluxes in the $0.5\mbox{--}2$\,keV and $2\mbox{--}10$\,keV bands are given in units of  $10^{-12}\,\mathrm{erg\,\mathrm{s}^{-1}\,\mathrm{cm}^{-2}}$}
\end{deluxetable*}
%
\subsection{\xmm\ Data Reduction}

\xmm\ \citep{2001Jansen} consists of the Optical Monitor (OM), the European Photon Imaging Camera (EPIC) with two EPIC-MOS CCDs and a single EPIC-pn CCD, and the Reflection Grating Spectrometers (RGS). Because of the higher effective area of EPIC-pn compared to EPIC-MOS and its consistency with EPIC-MOS data, we only consider EPIC-pn data in the $0.3\mbox{--}10.0$~keV energy band for the X-ray spectral analysis. The generation of updated calibration files (CIF) and the Observation Data Files (ODF) summary file follows the standard procedures by using the \xmm\ Science Analysis System (SAS 18.0.0). The EPIC-pn data are produced using \texttt{EPPROC} and processed with the standard filtering criterion. Then we remove periods of high background (total duration $\sim1.4\,$ks) in the light curve by creating a Good Time Interval (GTI) file using the task \texttt{TABGITGEN}. The source data are extracted from a circular region with a radius of 28 arcseconds centered on the source, and the background spectra are extracted from a nearby circular region with the same radius but excluding source photons. The influence of event pile-up has been checked by using the SAS task \texttt{EPATPLOT} and is found negligible during the observation. The Redistribution Matrix File (RMF) and Ancillary Region File (ARF) are created by using the SAS tasks \texttt{RMFGEN} and \texttt{ARFGEN} respectively. Finally, we group the spectral data to have a minimum of 30 counts per energy bin with task \texttt{GRPPHA}.

\subsection{\nustar\ Data Reduction}

The reduction of the \nustar\ \citep{2013Harrison} data was conducted following the standard procedures using the \nustar\ Data Analysis Software (NUSTARDAS v.1.8.0), and updated calibration files from \nustar\ CALDB v20190812. We produce calibrated and filtered event files with NUPIPELINE. The South Atlantic Anomaly (SAA) passages are calculated with strict filter \texttt{SAACALC=2} by using NUCALCSAA. We utilize the task package NUPRODUCTS to extract source spectra and light curves from a circular region of radius $82$ arcseconds centered on the source, as well as those of background from the a circular region of radius $145$ arcseconds free from source contamination. The spectra are grouped to at least 30 counts in each bin in order to have sufficiently high signal-to-noise ratio. The spectra of FPMA and FPMB are jointly analyzed between $3.0\mbox{--}79.0$\,keV in this paper.
%
\begin{figure*}[ht!]
\centering
\includegraphics[width=0.49\textwidth,trim={0 0.0cm 0 0}]{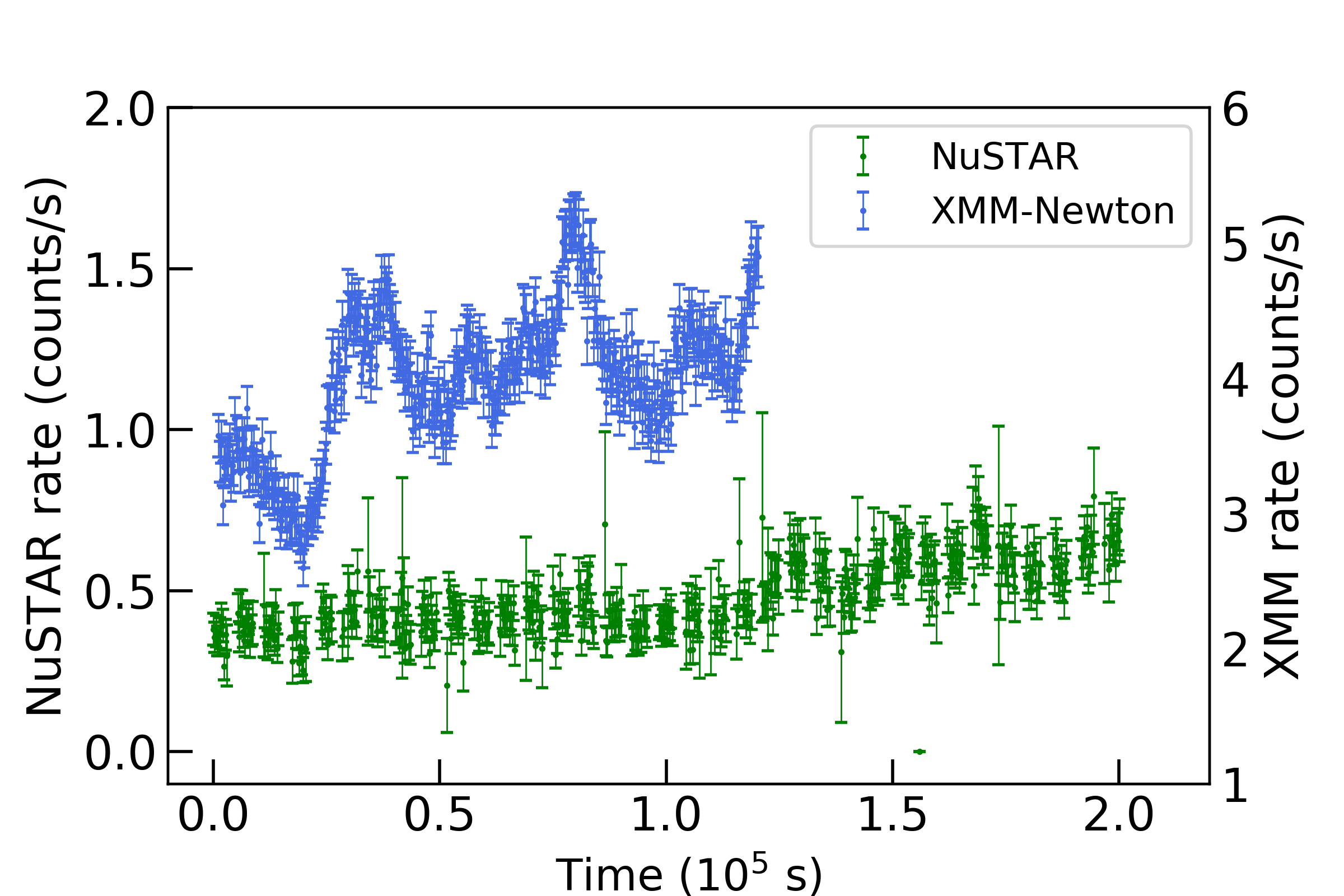}
\includegraphics[width=0.5\textwidth,trim={0 0.0cm 0 0}]{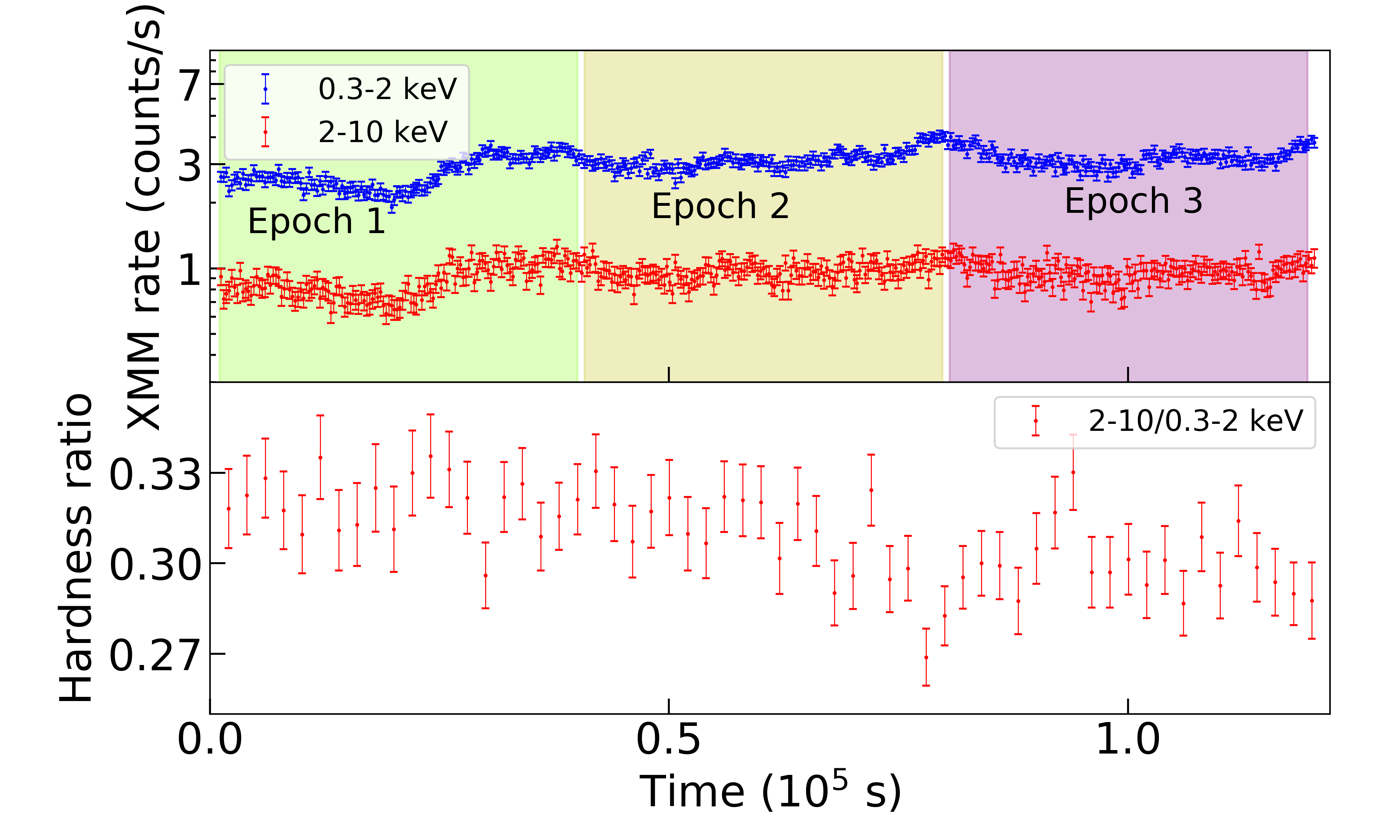}
\caption{
({\it left:}) Light curves of \nustar\ and simultaneous \xmm\ observations of \eso\ on September 24th of 2016 binned in 300\,s intervals . The net exposure times of \xmm\ and \nustar\ are $\sim 120$\,ks and $\sim 100$\,ks respectively. The \xmm\ light curve shows an obvious fluctuation while \nustar\ light curve is relatively moderate. ({\it right:}) Upper panel is the investigation of \xmm\ light curve in $0.3\mbox{--}2.0$\,keV and $2\mbox{--}10$\,keV bands. Three selected epochs are defined within green ({\it Epoch 1}), yellow ({\it Epoch 2}) and purple ({\it Epoch 3}) regions. Lower panel shows \xmm\ hardnesss ratios ($2\mbox{--}10/0.3\mbox{--}2.0$\,keV) binned in 2\,ks intervals for clarity to show the decreasing trend.
}
\label{fig:lightcurve}
\end{figure*}
%
\begin{figure}[ht!]
\centering
\includegraphics[width=0.49\textwidth,trim={0 0.0cm 0 0}]{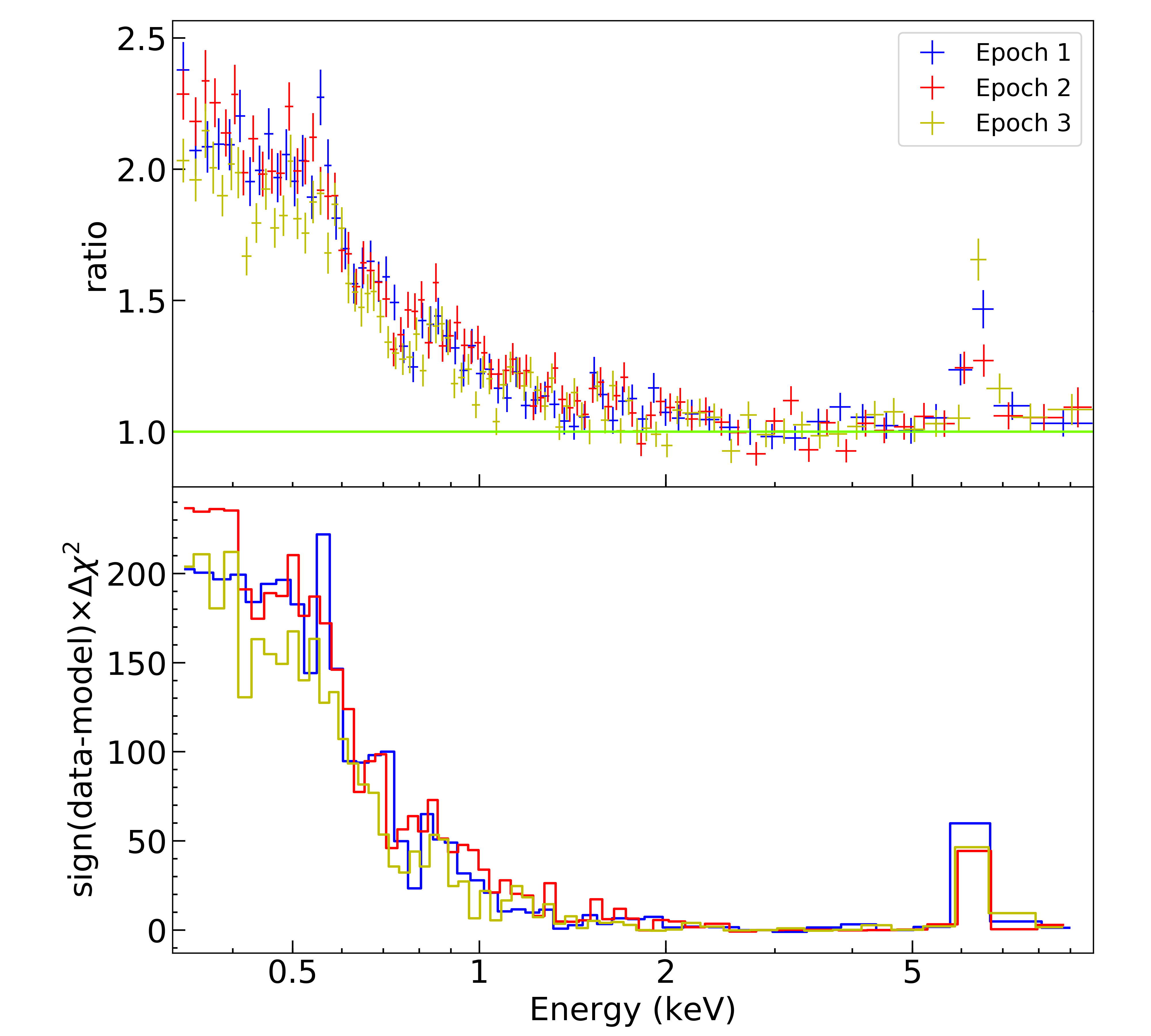}
\caption{
\xmm's Data-to-model ratio (upper panel) and the residuals (lower panel) with respect to a fit with a power-law model between $2\mbox{--}5$ and $8\mbox{--}10$\,keV (i.e. ignoring the soft excess and the iron complex region) for 3 different epochs marked in the upper right panel of Figure~\ref{fig:lightcurve}.
}
\label{fig:variability}
\end{figure}

\subsection{Light Curves and Variability}

The light curves of \nustar\ and simultaneous \xmm\ observations are presented in the left panel of Figure~\ref{fig:lightcurve}, binned in 300\,s intervals. The \nustar\ light curve shows a slight increase of the count rates (from $\sim$0.4 to $\sim$0.6\,counts/s) while significant fluctuations appear in the \xmm\ light curve. To investigate this variability, we extracted the \xmm\ light curves in the $0.3\mbox{--}2.0$\,keV and $2\mbox{--}10$\,keV bands, shown in the upper right panel of Figure~\ref{fig:lightcurve}, which shows that both soft and hard energy bands vary simultaneously. Moreover, we plot the \xmm\ hardness ($2\mbox{--}10/0.3\mbox{--}2.0$\,keV) ratio, binned in 2\,ks intervals for clarity in the lower right panel of Figure~\ref{fig:lightcurve}, presenting a decreasing trend. We have also extracted the spectra from the three equally divided epochs marked as the {\it green}, {\it yellow} and {\it purple} regions in the upper panel of Figure~\ref{fig:lightcurve}. A simple spectral fitting is performed to the spectra of the three epochs by using a power-law model between $2\mbox{--}5$\,keV and $8\mbox{--}10$\,keV (i.e. ignoring $<2\,$keV and $5\mbox{--}8$\,keV). Data-to-model ratios and $\Delta \chi^{2}$ are shown in Figure~\ref{fig:variability}. The residuals to the model show clear presences of a soft excess below $<2$\,keV and Fe K emission near $\sim6\mbox{--}7$\,keV. There is a hint that the soft excess is less variable than the power-law emission, owing to the slightly lower ratio in Epoch 3, which is the brightest epoch. Nonetheless, the spectral changes are very subtle, indicating the main variability originates from the entire broad band spectra.

\section{Spectral Analysis} \label{sec:spectra}
As mentioned above, since the spectrum only shows subtle variations during the observations, we focus on the time-averaged spectrum for the rest of the paper.
We analyze the time-averaged \nustar\ and \xmm\ spectra simultaneously using the {\sc xspec} (v12.10.1f) package \citep{1996Arnaud}. To account for the differences between the detector responses of FPMA/B and EPIC-pn, we include a cross-calibration factor, which is fixed to unity for the FPMA spectra, but varies freely for FPMB and EPIC-pn \citep{2015Madsen}. We employ the \csq\ statistics and estimate all parameter uncertainties at $90\%$ confidence level corresponding to $\Delta\chi^{2} = 2.71$. We include the absorption model \tbabs\ to describe the Galactic absorption, using the recommended photoelectric cross sections of \cite{1996Verner}, and solar abundances calculated of \citet{2000Wilms}. The galactic hydrogen column density is fixed at $N_\mathrm{H}^\mathrm{Gal}=1.75 \times 10^{20} \,\mathrm{cm^{-2}}$ \citep{2005Kalberla} during the spectral fitting. 

There has been a previous report of several soft X-ray narrow emission lines and a warm absorber affecting the low-energy flux in AG14. Although we do not present a full analysis here, we have checked the RGS spectrum and confirmed those lines around $0.55$\,keV, which are commonly detected in unobscured AGNs \citep[e.g.,][]{2016Reeves,2014Laha,2016Lohfink}. A series of emission lines corresponding to N~{\sc vii}, the O~{\sc vii} triplet, and O~{\sc viii} are detected, which can be modeled with five narrow \gaussian\ profiles, with the centroid energies fixed at the laboratory values, and with normalization values fixed at those reported by AG14. For the warm absorber, we model it by constructing ionized absorption grids with the {\sc xstar} photoionization code \citep{2001Kallman}, assuming solar abundances, non-turbulent velocity, and a $\Gamma=2$ power-law input spectrum. The column density $N_\mathrm{H}^\mathrm{ion}$ and the ionization parameter $\xi^\mathrm{ion}$ of the absorber are free to vary during our analysis. The fixed \gaussian\ models and the warm absorber model are applied to all the subsequent spectral fits.
%
\begin{figure}[ht!]
\centering
\includegraphics[width=0.49\textwidth,trim={0 0.0cm 0 0}]{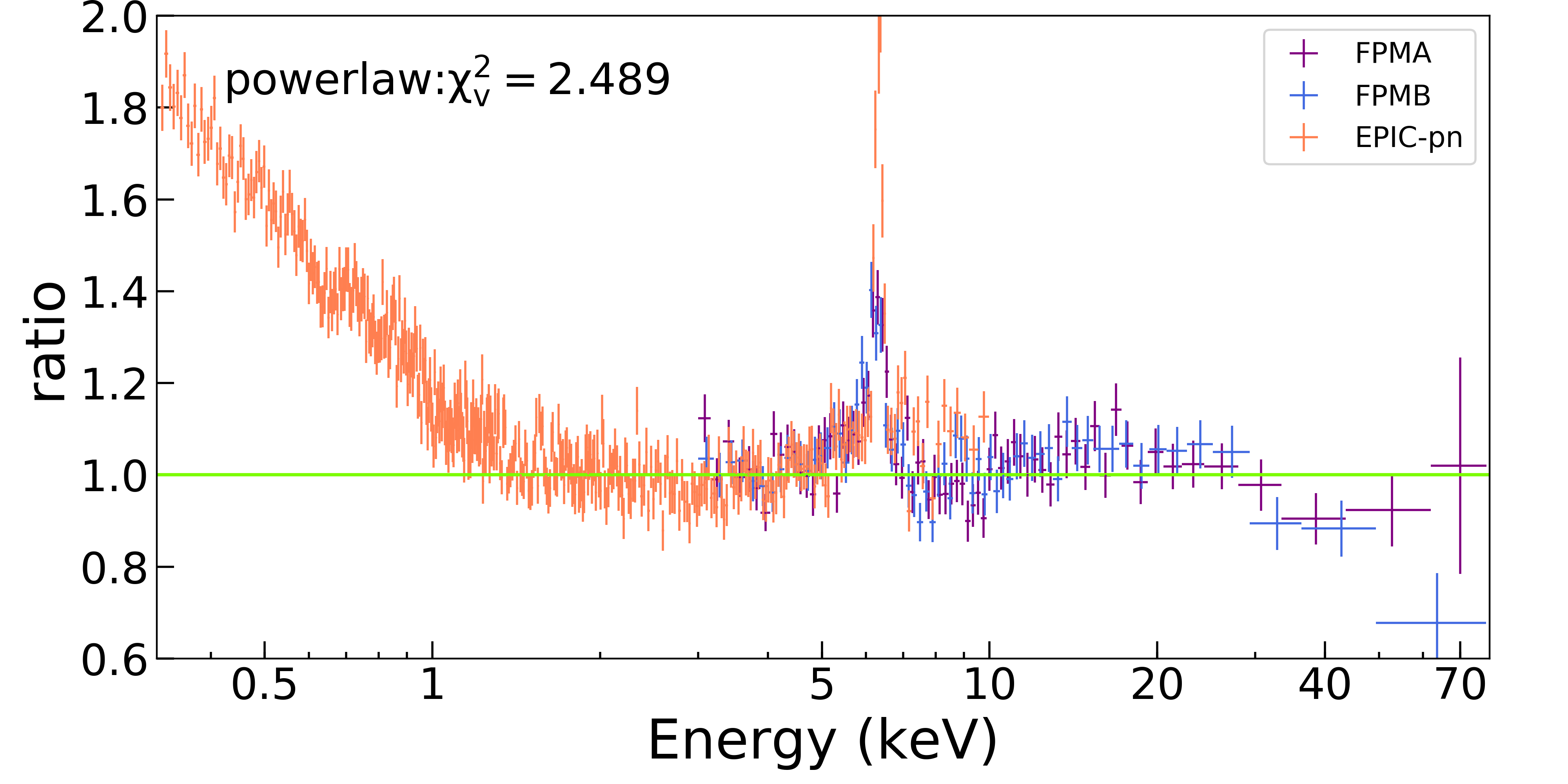}
\caption{
Data-to-model ratio of a simple power-law model (\pl), presenting the soft excess below 2\,keV, the iron K emission line at $\sim 6.4$\,keV, a (apparently weak) Compton hump peaking at $\sim 20$\,keV, and a cutoff at high energy. The ratio points are binned again to at least 20 counts for each bin (i.e., at least $30\times20$ counts for each bin in the plot) for clarity.
}
\label{fig:powerlaw}
\end{figure}

We start the fitting with a simple \pl\ model. Data-to-model ratios are shown in Figure~\ref{fig:powerlaw}. For clarity, the ratio points are binned again by a factor of 20 for each bin in the plot (i.e., each point represents at least $30\times20$ counts), which is the same in the following figures if not mentioned. The typical AGN spectral features can be seen: a soft excess below $\sim 2$\,keV, Fe K$\alpha$ emission at $\sim 6.4$\,keV, a Compton hump peaking at $\sim 20$\,keV, and a cutoff at high energy. Both the \xmm\ and \nustar\  spectra show a consistent shape for the iron K-shell emission line.

\begin{figure}[ht!]
\centering
\includegraphics[width=0.49\textwidth,trim={0 0.0cm 0 0}]{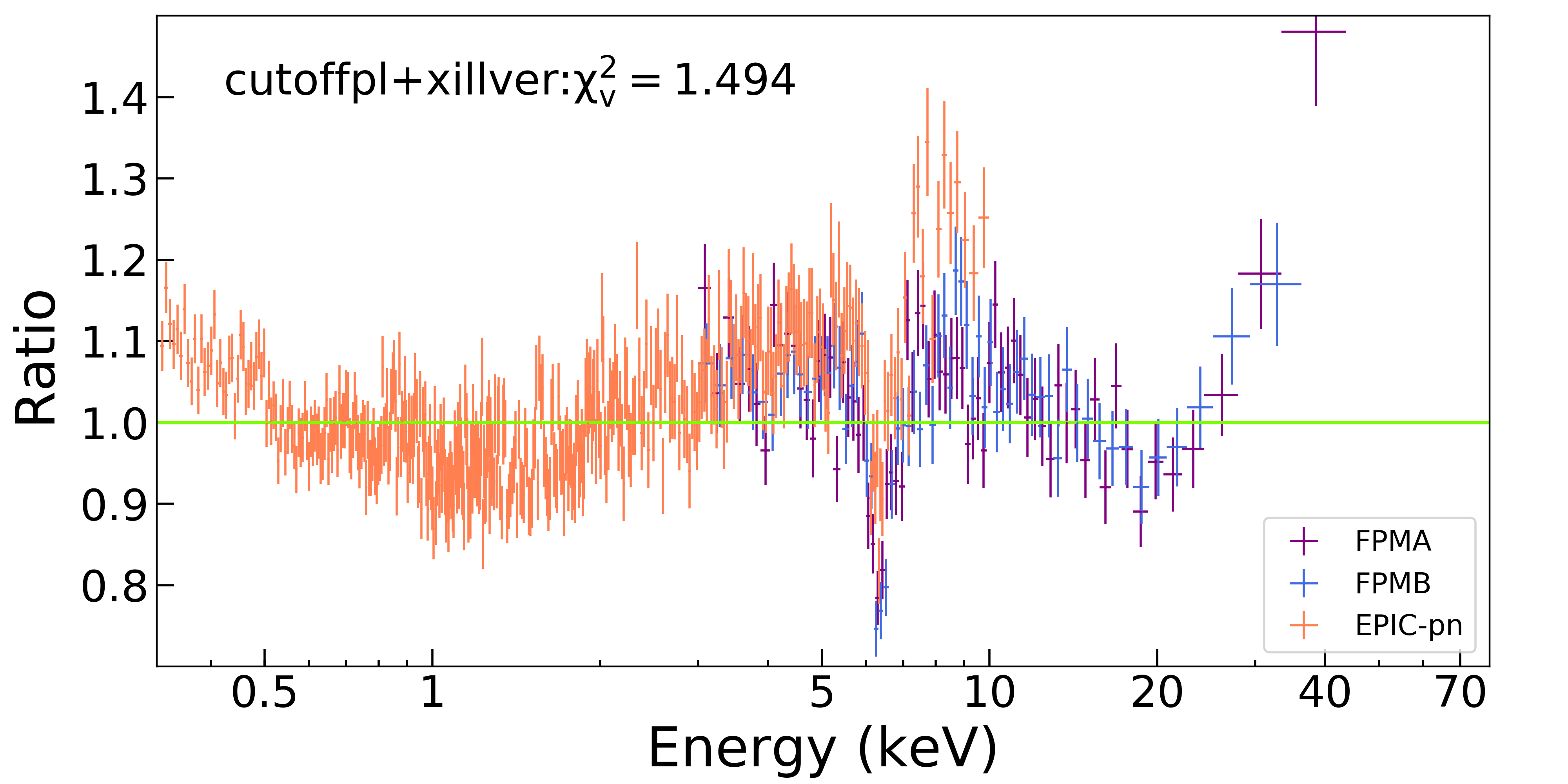}
\caption{
Data-to-model ratio of a simple power-law with a distant reflection model (\xillver).
}
\label{fig:xillver}
\end{figure}

These features can be partially accounted for by a reprocessing of X-ray photons in a neutral and distant material, free from relativistic effects, possibly at the broad line region \citep[e.g,][]{2016Nardini,2016Costantini}, or even the torus \citep[e.g.,][]{2007Yaqoob,2018Marinucci}. We thus fit the data with a cutoff power-law model (\cutoffpl) plus a neutral cold disk reflection model \citep[\xillver;][]{2010Garc,2011Garc,2013Garc}. Data-to-model ratios are shown in Figure~\ref{fig:xillver}. The photon index and energy cut-off are tied between these two components. In the \xillver\ model, the abundance of iron is assumed at the solar value (\afe$=1$) for simplicity. The ionization parameter, defined as the ratio of ionizing photon flux $F_\mathrm{X}$  ($0.1-1000$\,keV) to the gas density $n_e$ ($\xi = 4\pi F_\mathrm{X}/n_{e}$), is set to its minimum ($\xi=1$\,erg\,cm\,s$^{-1}$), since we do not expect the distant material to be highly ionized. The inclination angle of the accretion disk is fixed at the reference value $i = 53^{\circ}$ (AG14). As shown by the ratios, the reflection features and soft excess are not well described by the \xillver\ model. In fact, the Fe K emission is over-predicted by the model, likely due to the fact that a single neutral model is trying to account for all the other spectral features. In our subsequent analysis, we instead use the \borus\ model \citep{2019Balokovi} for the distant reflection, which also allows the reprocessing medium to be optically-thin.

%

Currently, the two competing models proposed to explain the soft excess are the warm corona and relativistic reflection. The fits of these two models to the \eso\ data are presented and discussed in the following sections. The different variants of models used in our fits are summarized in Table~\ref{tab:summodels}. The data-to-model ratios of the fits are depicted in the left column of Figure~\ref{fig:ratios}. The best-fit results are summarized in Table~\ref{tab:models}, where $\nu$ is the number of degrees of freedom (d.o.f) and $\chi_{\nu}^2=\chi^{2}/\nu$ is the reduced $\chi^{2}$. 

\begin{deluxetable}{cl}[ht!]
\tablecaption{Summary of the models used for the warm corona, the relativistic reflection and the hybrid hypotheses.
\label{tab:summodels}}
\tablecolumns{2}
\tablewidth{0pt}
\tablehead{
\colhead{Model} & \colhead{Components}
}
\startdata
{\it A} & \tbabs*\xstar*(\nthComp+\nthComp+\borus) \\
{\it B} & \tbabs*\xstar*(\nthComp+\relxilllpD+\borus) \\
{\it C} & \tbabs*\xstar*(\nthComp+\nthComp+\borus+\relxillCp) \\
\enddata
\end{deluxetable}
\subsection{{\it Model~A}: Warm Corona}\label{sec:testWC}

A warm ($T_{e} \sim 10^{5-6}$\,K) and optically thick ($\tau \sim 10\mbox{--}40$) corona model has been proposed to explain the observed soft excess in AGNs \citep[e.g.,][]{1998Magdziarz,2012Jin,2020Middei,2018Porquet,2018Petrucci,2019Ursini}. In this case, the thermal seed photons emitted from the accretion disk experience up-scattering via Comptonization in a warm corona. The soft excess is the high-energy tail of the resulting Comptonized spectrum. This corona is assumed to be an extended, slab-like plasma at the upper layer of disk, which is cooler than the hot ($T_e\sim 10^8-10^9$\,K) centrally-located and more compact corona responsible for the non-thermal power-law continuum.


\def\nh{$1.75$}	   
\def\zz{$0.012$}   
\def\seedkTbb{$3$}   
\def\wanh{$6$}
\def\walogxi{$1.4$}

\def\mIwanh{$4^{+1}_{-3}$}
\def\mIwalogxi{$1.4^{+0.3}_{-0.2}$}
\def\mIHCga{$1.58^{+0.01}_{-0.01}$}   
\def\mIHCkTe{$11.8^{+0.8}_{-0.7}$}   
\def\mIHCnor{$1.88^{+0.04}_{-0.04}$}  
\def\mIWCga{$2.5^{\star}$}   
\def\mIWCkTe{$0.19^{+0.01}_{-0.03}$}  
\def\mIWCnor{$1.9^{+0.2}_{-0.6}$}  
\def\mIincl{$53^{\star}$}            
\def\mIxillvernorm{$2.1^{+0.1}_{-0.1}$} 
\def\mIAfe{$>9.38$}       
\def\mICB{$1.00^{+0.01}_{-0.01}$} 
\def\mICE{$0.75^{+0.01}_{-0.01}$}   

\def\mIchi{$2411.70$}
\def\mIdof{$2341$}
\def\mIredc{$1.03020$}

\def\mIIwanh{$3^{+2}_{-2}$}
\def\mIIwalogxi{$1.4^{+0.2}_{-0.2}$}
\def\mIIHCga{$1.61^{+0.01}_{-0.01}$}   
\def\mIIHCkTe{$20^{+5}_{-3}$}   
\def\mIIHCnor{$1.85^{+0.05}_{-0.04}$}  
\def\mIIWCga{$2.5^{\star}$}   
\def\mIIWCkTe{$0.18^{+0.02}_{-0.03}$}  
\def\mIIWCnor{$1.6^{+0.5}_{-0.6}$}  
\def\mIIBonh{$23.2^{+0.3}_{-0.2}$}   
\def\mIIBocf{$>0.36$}  
\def\mIIAfe{$1.0^{+0.3}_{-0.2}$} 
\def\mIIBonorm{$0.5^{+0.1}_{-0.1}$} 
\def\mIICB{$1.00^{+0.01}_{-0.01}$} 
\def\mIICE{$0.76^{+0.01}_{-0.01}$}   
\def\mIIchi{$2359.77$}
\def\mIIdof{$2339$}
\def\mIIredc{$1.00888$}

\def\mVwanh{$5^{+2}_{-3}$}
\def\mVwalogxi{$1.4^{+0.3}_{-0.5}$}
\def\mVHCga{$1.72^{+0.02}_{-0.02}$}   
\def\mVHCkTe{$35^{+18}_{-11}$}   
\def\mVHCnor{$1.99^{+0.07}_{-0.03}$}  
\def\mVWCga{$2.5^{\star}$}   
\def\mVWCkTe{$0.08^{+0.03}_{-0.01}$}  
\def\mVWCnor{$0.03^{+0.19}_{-0.02}$}  
\def\mVBonh{$23.0^{+0.1}_{-0.1}$}   
\def\mVBocf{$>0.8$}  
\def\mVAfe{$0.5^{+0.1}_{-0.1}$} 
\def\mVBonorm{$1.1^{+0.1}_{-0.2}$} 
\def\mVqI{$3^\star$}    
\def\mVqII{$3^{\star}$}    
\def\mVRbr{$100^{\star}$}    
\def\mVa{$0.998^\star$}                   
\def\mVi{$<14$}       
\def\mVRin{$50^\star$}       
\def\mVlxi{$1.3^{+0.1}_{-0.1}$}   
\def\mVnor{$0.09^{+0.02}_{-0.03}$}   
\def\mVCB{$1.00^{+0.01}_{-0.01}$} 
\def\mVCE{$0.76^{+0.01}_{-0.01}$}   
\def\mVchi{$2341.14$}
\def\mVdof{$2336$}
\def\mVredc{$1.00220$}

\def\mIIIwanh{$2^{+1}_{-1}$}
\def\mIIIwalogxi{$1.4^{+0.2}_{-0.3}$}
\def\mIIIHCga{$1.75^{+0.02}_{-0.01}$}               
\def\mIIIHCkTe{$>78$}           
\def\mIIIHCnor{$1.94^{+0.03}_{-0.05}$}        
\def\mIIIqI{$3.9^{+0.4}_{-0.5}$}    
\def\mIIIqII{$3^{\star}$}    
\def\mIIIRbr{$10^{\star}$}    
\def\mIIIa{$>0.924$}                   
\def\mIIIi{$37^{+10}_{-21}$}       
\def\mIIIlxi{$0.4^{+0.1}_{-0.1}$}   
\def\mIIIlne{$>18.2$}                     
\def\mIIInor{$0.23^{+0.05}_{-0.01}$}   

\def\mIIIBonh{$22.8^{+0.3}_{-0.2}$}   
\def\mIIIBocf{$>0.5$}  
\def\mIIIAfe{$1.7^{+0.8}_{-0.1}$} 
\def\mIIIBonorm{$6^{+3}_{-3}$} 
\def\mIIICB{$1.00^{+0.01}_{-0.01}$} 
\def\mIIICE{$0.76^{+0.01}_{-0.01}$}   
\def\mIIIchi{$2375.48$}
\def\mIIIdof{$2335$}
\def\mIIIredc{$1.01734$}
\def\mIVwanh{$4^{+1}_{-1}$}
\def\mIVwalogxi{$1.4^{+0.1}_{-0.1}$}
\def\mIVHCga{$1.74^{+0.02}_{-0.02}$}               
\def\mIVHCkTe{$>51$}           
\def\mIVHCnor{$2.02^{+0.02}_{-0.05}$}      

\def\mIVh{$3.0^{+1.0}_{-0.3}$}   
\def\mIVa{$>0.927$}                   
\def\mIVi{$32^{+4}_{-5}$}       
\def\mIVlxi{$0.4^{+0.3}_{-0.1}$}   
\def\mIVlne{$>18.3$}                     
\def\mIVnor{$4.8^{+0.8}_{-0.9}$}   
\def\mIVBonh{$22.8^{+0.2}_{-0.2}$}   
\def\mIVBocf{$>0.8$}  
\def\mIVAfe{$1.5^{+0.4}_{-0.3}$} 
\def\mIVBonorm{$0.7^{+0.1}_{-0.1}$} 
\def\mIVCB{$1.00^{+0.01}_{-0.01}$} 
\def\mIVCE{$0.76^{+0.01}_{-0.01}$}   
\def\mIVchi{$2376.27$}
\def\mIVdof{$2335$}
\def\mIVredc{$1.01768$}

\begin{deluxetable*}{lcccccccc}[ht!]
\tablecaption{Best-fit parameters and fitting statistics of the warm corona (A), relativistic reflection (B) and the hybrid (C) scenarios.
\label{tab:models}}
\tablecolumns{6}
\tablewidth{0pt}
\tablehead{
\colhead{Description} & \colhead{Component} & \colhead{Parameter} &
\colhead{Model~A} & \colhead{Model~B}& \colhead{Model~C} 
}
\startdata
Galactic Absorption & {\tt tbabs}         & $N_\mathrm{H}^\mathrm{Gal}$ ($10^{20}$ cm$^{-2}$)& $\nh^{\star}$       &  $\nh^{\star}$ &$\nh^{\star}$   \\
\hline
Warm Absorption & \xstar\       & $N_\mathrm{H}^\mathrm{ion}$ ($10^{20}$ cm$^{-2}$)& \mIIwanh\        &\mIVwanh\  & \mVwanh\  \\
& \xstar\         & $\log{[\xi^\mathrm{ion}/\mathrm{erg}\,\mathrm{cm}\,\mathrm{s}^{-1}]}$ & \mIIwalogxi\      & \mIVwalogxi\ & \mVwalogxi\   \\
\hline
Hot Corona          & {\tt nthComp}       & $\Gamma^\mathrm{HC}$               & \mIIHCga\   &  \mIVHCga\ &\mVHCga\  \\
                    & {\tt nthComp}       & $kT_\mathrm{e}^\mathrm{HC}$ (keV)   & \mIIHCkTe\  & \mIVHCkTe & \mVHCkTe\  \\
                    & {\tt nthComp}       & $N_\mathrm{HC}$ ($10^{-3}$)         & \mIIHCnor\  & \mIVHCnor\ &\mVHCnor\  \\
\hline
Warm Corona         & {\tt nthComp}       & $\Gamma^\mathrm{WC}$    & \mIIWCga\ &  \nodata\ &  \mVWCga\   \\
 					& {\tt nthComp}       & $kT_\mathrm{e}^\mathrm{WC}$ (keV)   & \mIIWCkTe\  & \nodata\ & \mVWCkTe\  \\
                    & {\tt nthComp}       & $N_\mathrm{WC}$ ($10^{-4}$)         & \mIIWCnor\   &\nodata\ & \mVWCnor \\
\hline
Relativistic Reflection & {\tt relxillCp}& $q$ &\nodata\ &\nodata\ & \mVqI\  \\
                        & {\tt relxilllpD} & $h$ ($R_\mathrm{Horizon}$) &\nodata\ &\mIVh & \nodata\     \\
                        & {\tt relxillCp/lpD}    & $a_\star$ ($cJ/GM^2$)                   &\nodata\ &\mIVa\  & \mVa\     \\
                        & {\tt relxillCp/lpD}    & $i$ (deg)   &\nodata\ &  \mIVi\  & \mVi\     \\
                        & {\tt relxillCp/lpD}    & $R_\mathrm{in}$ (ISCO)   &\nodata\ & $1^\star$          & \mVRin\  \\
                        & {\tt relxillCp/lpD}    & \logxi\   &\nodata\ & \mIVlxi\  & \mVlxi\    \\
						& {\tt relxilllpD}    & $\log{[n_\mathrm{e}/\mathrm{cm}^{-3}]}$    &\nodata\ & \mIVlne\  & \nodata\   \\
					    & {\tt relxillCp/lpD}    & $N_\mathrm{r}$ ($10^{-4}$) &\nodata\ & \mIVnor\ & \mVnor\   \\
\hline
Neutral Reflection  & {\tt Borus12}       & $\log{[N_\mathrm{H,tor}/\mathrm{cm}^{-2}]}$ & \mIIBonh\ & \mIVBonh\ & \mVBonh\   \\
                    & {\tt Borus12}       & $C_\mathrm{tor}$    & \mIIBocf\ & \mIVBocf\ & \mVBocf\ \\
                    & {\tt Borus12}       & $A_{\mathrm{Fe}}$ & \mIIAfe\ & \mIVAfe\ & \mVAfe\   \\
                    & {\tt Borus12}       & $N_\mathrm{B}$ ($10^{-2}$) & \mIIBonorm\ & \mIVBonorm\ & \mVBonorm\  \\
\hline
Cross-normalization & \nustar\ FPMB       & $C_\mathrm{FPMB}$        & \mIICB\ & \mIVCB\ & \mVCB\            \\
                    & \xmm\ EPIC-pn        & $C_\mathrm{EPIC-pn}$    & \mIICE\ & \mIVCE\ & \mVCE\     \\
\hline
                    && $\chi^2$             & \mIIchi\  & \mIVchi\ & \mVchi\         \\
                    && $\nu$ (d.o.f)        & \mIIdof\  & \mIVdof\ & \mVdof\            \\
                    && $\chi_{\nu}^2$      & \mIIredc\   & \mIVredc\ & \mVredc\   \\
\enddata
\tablenotetext{\star}{The parameter is pegged at this value.}
\end{deluxetable*}

We now implement a more physically motivated model, \nthComp\ \citep{1996Zdziarski,1999zycki}, to represent the Comptonization plasmas. Two \nthComp\ components are employed to describe both the warm and hot coronae. The hot \nthComp\ reproduces a power-law continuum emitted by the Comptonized photons in a central hot electron gas (previously modeled with \cutoffpl). A second \nthComp\ model is used for the warm corona component, which  is characterized by the continuum slope, $\Gamma$, the temperature of the covering electron gas, $kT_\mathrm{e}$, and the seed photon temperature, $kT_{bb}$. We assume the same disk-blackbody seed photon \citep[e.g.,][]{1984Mitsuda,1986Makishima} temperature for both components, $kT_\mathrm{bb}=3$\,eV, which is the typical temperature for the accretion disks of AGN. We leave the electron temperature $kT_\mathrm{e}$ and normalization variable in \nthComp. The spectral slope of the hot corona $\Gamma^\mathrm{HC}$ is also variable, but that of the warm corona is fixed at $\Gamma^\mathrm{WC}=2.5$. To model the reflection spectrum, we exploit a physical torus model \borus\ \citep{2019Balokovi}.

%
\begin{figure}[ht!]
\centering
\includegraphics[width=0.5\textwidth,trim={0 0.0cm 0 0}]{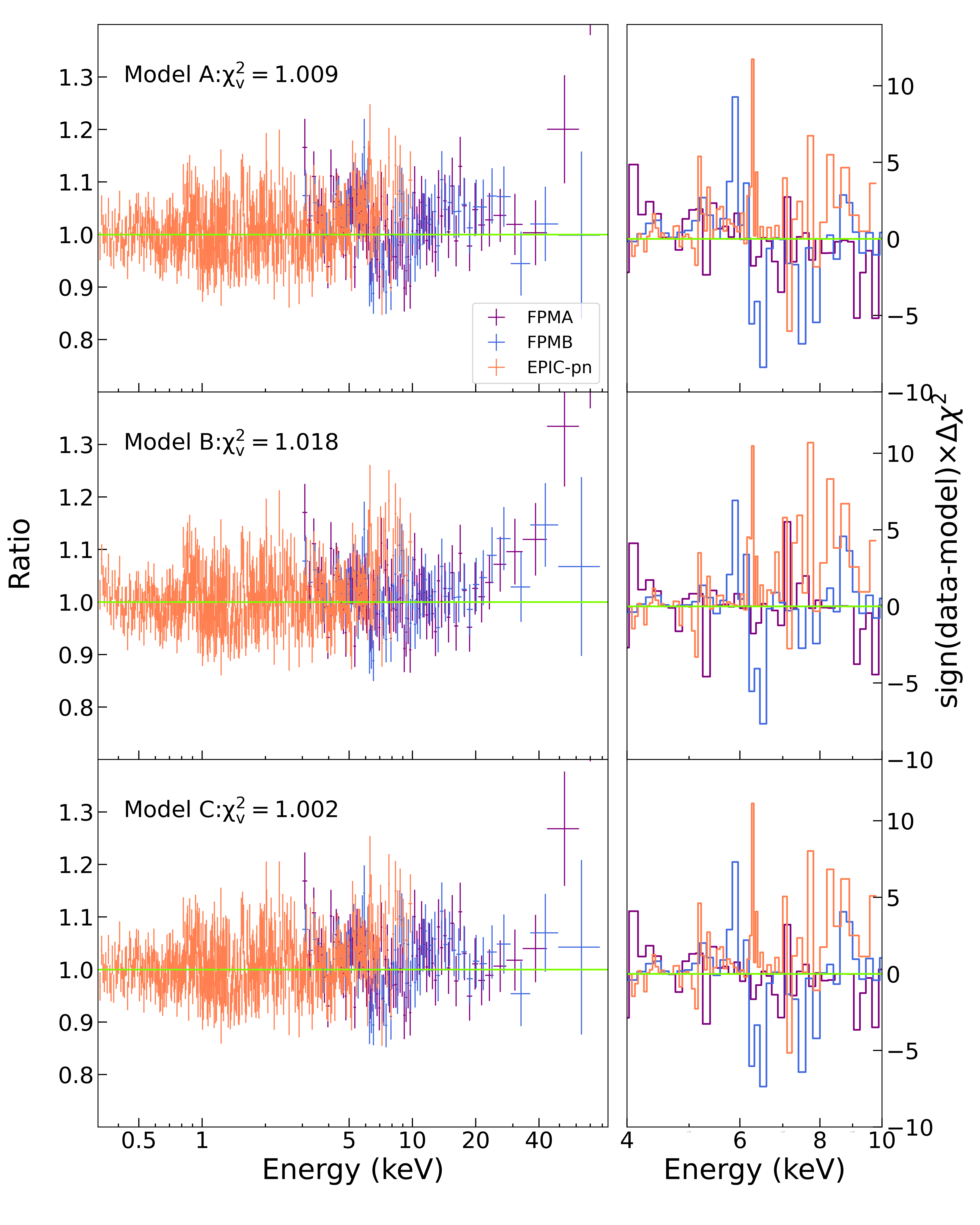}
\caption{
({\it Left}:) Data-to-model ratios for the fitting of the warm corona ({\it Model~A}) and relativistic reflection ({\it Model~B}) scenarios with both hot corona and distant reflection component, and a hybrid model of {\it Model~C}. The instrument data, model components, and the statistical results are shown in panels. ({\it Right}:) The zoom-in residuals with respect to fits identical to the left model between $4\mbox{--}10$\,keV.
}
\label{fig:ratios}
\end{figure}
The \borus\ model calculates the transmitted and scattered X-ray photons in the cold, neutral and uniformly distributed gas with a toroidal geometry. The primary power-law X-ray continuum is emitted from the Comptonized geometrically central source, described by \nthComp. The torus covering factor, $C_\mathrm{tor}$, is defined as cosine of $\theta_{tor}$, which is the half-opening angle of the polar cutouts, measured from the symmetry axis toward the equator. Moreover, the \borus\ model allows us to calculate the average column density of the torus ($N_\mathrm{H,tor}$) since the model only includes the reprocessed component, and the line-of-sight component ($N_\mathrm{H,los}$) could be added separately in {\sc xspec}. In our spectral analysis, due to the relatively poor constraint on the cosine of torus inclination angle, we fixed it to $\cos{\theta_\mathrm{inc}}=0.8$ ($\theta_\mathrm{inc}=37^\circ$). The value is determined by optical imaging \citep{2018Humire}. The iron abundance, covering factor, average column density of torus and normalization are left free in our fits. The photon index and the electron gas temperature are linked to those of the hot corona. 

The warm corona model results in excellent fit statistic $\chi^2_{\nu}\sim1.009$ (see the fourth column of Table~\ref{tab:models}) and models the residuals in Figure~\ref{fig:xillver} well (the upper left panel of Figure~\ref{fig:ratios}). The fitting requires a reasonable iron abundance $A_\mathrm{Fe}=1.0^{+0.3}_{-0.2}$. Both the photon index and the hot gas temperature have relatively low values ($\Gamma^\mathrm{HC}=1.61^{+0.01}_{-0.01}$, $kT_\mathrm{e}^\mathrm{HC}=20^{+5}_{-3}$\,keV). The electron temperature of the warm corona is $kT_\mathrm{e}^\mathrm{WC}=0.18^{+0.02}_{-0.03}$\,keV, which is consistent with the range ($0.1\mbox{--}1$\,keV) reported by \citet{2018Petrucci}. The calculated average column density of the torus is $\log{[N_\mathrm{H,tor}/\mathrm{cm}^{-2}]}=23.2^{+0.3}_{-0.2}$ and the covering factor is $C_{tor}>0.36$, corresponding to $\theta_{tor}<69^\circ$.

{\it Model~A} proves that the warm corona scenario is able to model the spectrum well from a statistical point of view. From the physical point of view, \borus\ describes a 3-dimensional torus rather than a simple 1-dimensional plane-like slab (as assumed in the \xillver\ model), indicating the geometrical considerations are important for this source. Moreover, compared with previous fits of a single power-law plus a \xillver\ component, {\it Model~A} describes the iron complex region with a neutral distant reflection model, although we still can see some systematic residuals in the upper left panel of Figure~\ref{fig:ratios} around the Fe K energies. However, the most peculiar aspect of the warm corona model is the relatively low temperature of the hot corona, which is uncommon for AGN \citep[e.g.,][]{1994Nandra,2017Ricci,2020Balokovi}. This is further discussed in Section~\ref{sec:discussWC}.

\subsection{{\it Model~B}: Relativistic Reflection}\label{sec:testRR}
The other popular explanation of the soft excess is that it is a part of the relativistic reflection spectrum from the inner accretion disk \citep[e.g.,][]{2006Crummy,2009Fabian,2013Walton,2018Jiang,2019Garc}. If the X-ray reprocessed photons are radiated on the inner region of the accretion disk, due to the relativistic effects from the supermassive black hole, the fluorescent features will be blurred, smoothing the entire reflection spectrum \citep{2005Fabian}. Moreover, it has recently been shown that the existence of the enhanced inner-disk density, above the commonly assumed value of $n_{e}=10^{15}\,\mathrm{cm}^{-3}$ \citep[e.g.,][]{1993Ross,2005Ross,2011Garc}, results in increased emission at soft energies ($\lesssim2$\,keV). This occurs because at high densities free-free heating (bremsstrahlung) in the disk atmosphere becomes dominant and results in an increased gas temperature \citep{2016Garc,2019Jiang}. As a consequence, relativistic reflection from a highly dense disk can lead to increased low-energy emission, which may account for the soft excess.

To test this model, we employ a relativistic reflection model with variable disk density to fit the data. Currently, the most advanced relativistic reflection model is \relxill\ \citep{2014Garc,2014Dauser}. It is a combination of \xillver\ \citep{2010Garc,2013Garc}, describing the disk reflection component, and a relativistic convolution code {\tt relconv} \citep{2010Dauser,2013Dauser}. Therefore, it provides an effective model to study the ionized reflection in strong gravitational fields. Specifically, we implement the relativistic reflection model \relxilllpD, a version of \relxill\ that allows for variable disk density, while assuming a high-energy cutoff fixed at $300$\,keV  \citep{2016Garc}. For the neutral distant reflection we still use \borus, since the geometrical aspects described by this model seem to be relevant in this source. 

The \relxilllpD\ \citep{2013Dauser,2014Garc,2016Garc} model is a flavor of \relxill\ that assumes a lamppost coronal geometry. We can estimate the distance between the corona and the black hole through this model. For the spectral slope, the hot corona, relativistic disk reflection and distant reflection components are tied together. The electron temperature of \borus\ is linked to that in the hot corona as well. The inner radius of the accretion disk $R_\mathrm{in}$ is set at the innermost stable circular orbit (ISCO) and the outer disk radius $R_\mathrm{out}$ is fixed at its default value ($R_\mathrm{out} = 400\,R_{\rm g}$), far from the inner regions of disk. The ionization parameter $\xi$, the hot corona temperature $kT_\mathrm{e}^\mathrm{HC}$ and the disk density $n_\mathrm{e}$ vary freely in \relxilllpD\ and the iron abundance is linked to that of \borus. 

The best-fit values and ratios are listed in the fifth column of Table~\ref{tab:models} and the middle left panel of Figure~\ref{fig:ratios}. The relativistic reflection picture provides a satisfactory fit ($\chi^2_\nu=1.018$), although it is slightly worse than the warm corona model, with larger residuals at high energies. This may be related to the \cutoffpl\ continuum assumed by \relxilllpD, as even for fairly high values of $E_{\rm{cut}}$ such a continuum constantly bends at all energies, while a physical Comptonisation model is more powerlaw-like below its high-energy cutoff (e.g. \citealt{2001Petrucci}); a version of \relxilllpD\ that allows for both a Comptonisation continuum and a variable electron temperature will be required to test this. We have also performed a fit where we fix $kT_{\rm{e}}$ to 100\,keV for the hot corona, broadly equivalent to $E_{\rm{cut}} = 300$\,keV (\citealt{2001Petrucci}), so that the treatment of the high-energy cutoff is more self-consistent across the different model components. This gives essentially identical results to those reported below, implying that the key reflection results are relatively insensitive to this issue.

The spectral slope found for model B is $\Gamma=1.74^{+0.02}_{-0.02}$ and the temperature of the hot corona only provides the lower limit $kT_\mathrm{e}^\mathrm{HC}>51\,\mathrm{keV}$. It requires a rapidly rotating black hole ($a_\star>0.927$) and the disk inclination angle is $i=28^{+3}_{-3}$\,deg. The height of the corona is $h=3.0^{+1.0}_{-0.3}$\,$R_\mathrm{Horizon}$ away from the black hole, where $R_\mathrm{Horizon}$ is the vertical event horizon of the Kerr black hole. In the relativistic reflection scenario, the required gas density is high (\logne$>18.3$), and the ionization is low ($\log{[\xi/\mathrm{erg}\,\mathrm{cm}\,\mathrm{s}^{-1}]}=0.4^{+0.1}_{-0.1}$). We find that the parameters of the distant reflection component are similar to those in the warm corona model. 

%
\begin{figure}[ht!]
\centering
\includegraphics[width=1\textwidth,width=0.49\textwidth,trim={0 0.0cm 0 0}]{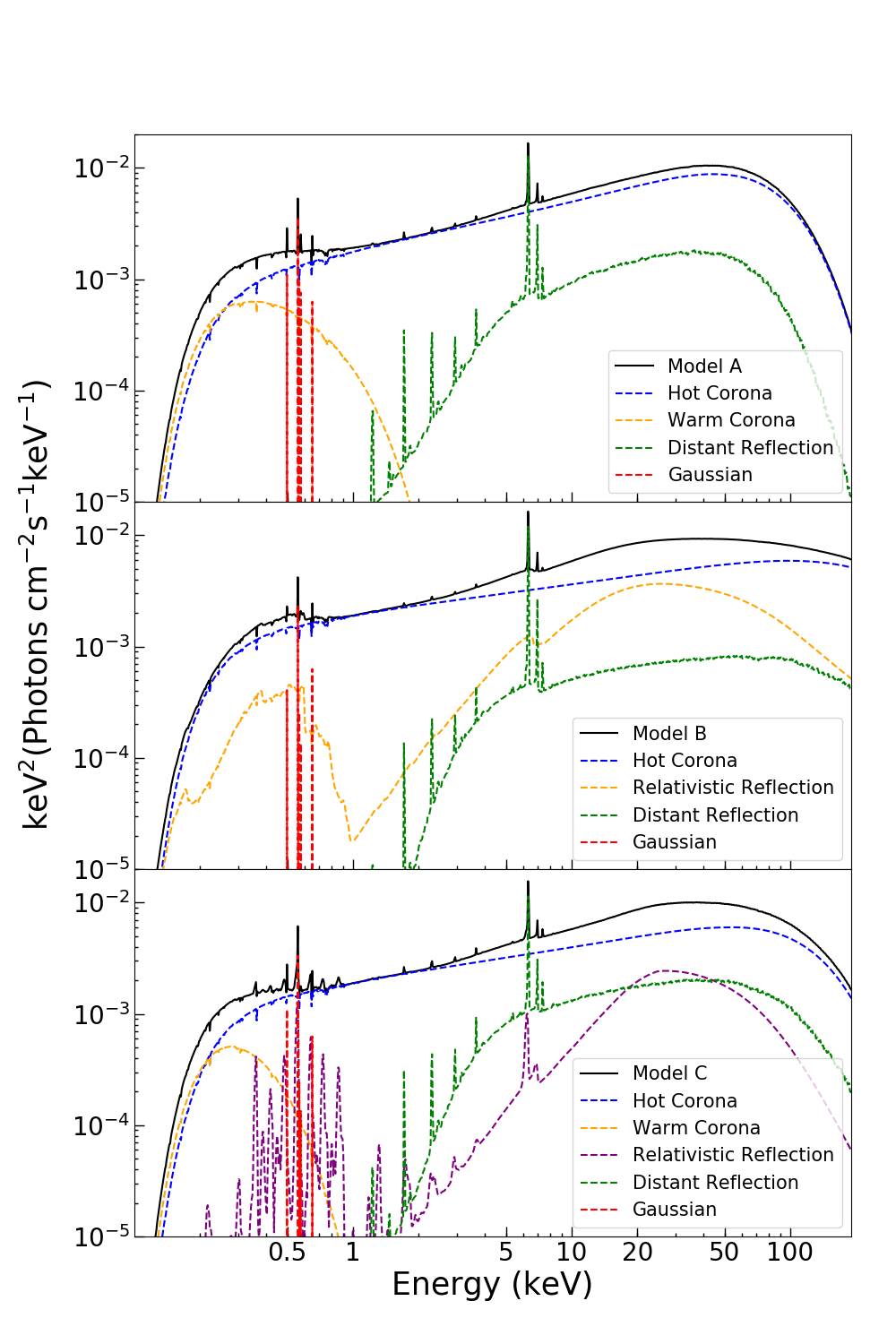}
\caption{
Model components of the best model fit in the warm corona scenario ({\it Model~A}, {\it Top}), the relativistic reflection scenario ({\it Model~B}, {\it Middle}), and the hybrid model ({\it Model~C}, {\it Bottom}). The {\it black} solid line is the total contribution of the model components. The {\it blue} and {\it green} dash lines are the primary X-ray continuum emitted by the hot corona and the reflection off the distant cold materials respectively. The {\it orange} dash line in the {\it top} and {\it bottom} panel is the warm corona component while that in the {\it middle} panel is the high-density relativistic reflection. The {\it purple} dash line in the {\it bottom} panel is the relativistic reflection component with a fixed density at $10^{15}\,\mathrm{cm}^{-3}$. The {\it red} dash lines are five emission lines reported in AG14.
}
\label{fig:eemods}
\end{figure}

Compared to the warm corona explanation, the relativistic reflection model is slightly worse in a statistical sense, although in general both models provide similarly good fits to the data. While in {\it Model~B} the covering factor ($C_\mathrm{tor}>0.8$) is high, the column density is low, so there is not much of a Compton hump coming from this distant reprocessor (see in Figure~\ref{fig:eemods}). The contributions from different components to {\it Model~A} and {\it Model~B} are plotted in the upper two panels of  Figure~\ref{fig:eemods}. Both fits provide a strong excess at low energies, either from the warm Comptonization or the blurred emission lines stimulated by hot coronal irradiation of a high-density disk. In {\it Model~B}, the relativistic reflection component shows a broad iron line and it models the systematic signals around $6.4\,$keV. When the data outside of $4\mbox{--}10\,$keV is ignored, the remaining statistics are $\chi^2=956~\mathrm{and}~940$ for {\it Models A and B}, respectively (see the right column of Figure~\ref{fig:ratios}). It reveals that the relativistic effect is strong for this source, which matches with the high spin value. The obvious difference between these two models (A and B) is shown in Figure~\ref{fig:eemods}. The relativistic reflection component in {\it Model B} contributes more photons than the distant reflection to the high energies. This explains the softer primary continuum of {\it Model~B} and makes the column density of the torus slightly lower, because the \borus\ model does not need to provide many high-energy photons.

\subsection{{\it Model~C}: Hybrid Model}
Although this is often not formally considered in detail in the literature, in the warm corona interpretation there should still be reflection from the standard disk outside of the warm corona. We therefore also test a hybrid model that includes a warm Comptonization component and a relativistic reflection component, both of which can contribute to the soft excess. Here we make use of the \relxillCp\ model, which assumes both a Comptonization continuum and a fixed density of $10^{15}\,\mathrm{cm}^{-3}$. We choose this model because at the radii outside of the warm corona any relativistic effects associated with the regions close to the black hole become negligible, and the emissivity index asymptotes towards $q=3$ for any reasonable lamppost height \citep[e.g,][]{2011Wilkins}. When fitting the warm corona model to a sample of nearby AGN, \citet{2012Jinb} found that the average radius of the warm corona is $\sim50\,R_\mathrm{g}$ (and is always $>10\,R\mathrm{g}$). We therefore assume that the emissivity profile follows a $q=3$ power law here. Since the inner radius of the standard disc is no longer the ISCO (as this is set by the radius of the warm corona; see further discussion in Section~\ref{sec:discussWC}), we also fix the spin to the maximal value ($a_\star=0.998$, compatible with both our results and also those of AG14) and instead initially allow the inner radius to vary. However, we find that when it is free to vary the inner radius is poorly constrained in this model, and so we present the results for $R_\mathrm{in}=50\,R_\mathrm{g}$ \citep[based on the results of][]{2012Jinb}. The electron temperature and spectral slope are tied with those of the hot corona. The inclination angle of the disk and the ionization parameter are left free to vary. 

The inclusion of the relativistic reflection component slightly reduces the fit statistics to $\chi^2_{\nu}\sim1.002$ (see the sixth column of Table~\ref{tab:models}) and faintly improves the remaining residuals seen with the pure warm corona model in the iron complex region (the right column of Figure~\ref{fig:ratios}). The contribution from different components is also plotted in the bottom panel of Figure~\ref{fig:eemods}. Compared to {\it Model~A}, the main changes are a softer X-ray continuum ($\Gamma^\mathrm{HC}=1.72^{+0.02}_{-0.02}$), and a higher temperature of the hot corona ($kT^\mathrm{HC}_\mathrm{e}=35^{+18}_{-11}$\,keV), a cooler warm corona ($kT_\mathrm{e}^\mathrm{WC}=0.08^{+0.03}_{-0.01}$\,keV) and the much lower normalization of the warm corona ($N_\mathrm{WC}=0.03^{+0.19}_{-0.02}\times10^{-4}$). The noteworthy parameters, compared with {\it Model~B}, are a nearly face-on inclination angle ($i<14^\circ$), a higher ionization state of the relativistic reflection component ($\log{[\xi/\mathrm{erg}\,\mathrm{cm}\,\mathrm{s}^{-1}]}\sim1.3$). We also calculate the unobscured flux of the warm corona and the relativistic reflection below 2\,keV, which are $F^\mathrm{WC}_\mathrm{<2\,keV}=3.31^{+0.16}_{-0.16}\times10^{-12}\,$\fluxunits, and 
$F^\mathrm{RR}_\mathrm{<2\,keV}=3.38^{+0.14}_{-0.14}\times10^{-13}\,$\fluxunits\ respectively, implying the bulk of the soft excess arises from the warm corona in the hybrid model.

This attempt seems to be a intermediate solution between the pure warm corona and relativistic reflection models. For example, the best-fit values of the hot corona temperature and the column density of the torus are the modest values and compatible with both of {\it Model~A} and {\it Model~B}. The co-existence of the two components also imposes the weakening on the normalization of both components, while the spectral slope is only compatible with the {\it Model~B} and is not consistent with the pure warm corona, although the the warm corona provides more photons for the soft excess. Therefore, we tend to consider the hybrid model to be a reference for our discussion, since the aim of this paper is to discuss the dominant mechanism of the soft excess.

\subsection{Testing the inclusion of UV data}
In order to investigate whether the UV data can provide any further insight in this case, we also extract the data from the Optical Monitor \citep[][OM]{2001Mason}, following the standard procedures \citep[e.g.,][]{2011Mehdipour,2018Petrucci}. The OM is an optical/UV instrument on board of \xmm\ which provides strictly simultaneous UV observations. Only imaging mode data through the UVM2 (2310 \AA) and UVW2 (2120 \AA) filters are available for this observation. They are processed and corrected using the standard {\tt omichain} pipeline with {\tt nsigma=3}, which takes all necessary calibration processes (e.g. source identification) into account. The output file is a combination of the source list from separate exposures on the object.

We performed additional fits with each of our models after including the available OM photometry. The optical extinction was found to be variable by \citet{2017agis}, but in the 2016 data (which is closest in time to our X-ray observations) was found to be $E_\mathrm{B-V}=0.57\pm0.07$, and is modeled by {\tt redden} in {\sc xspec}. We also include a $5\%$ systematic error on the OM data in order to account for any cross-normalisation issues between the OM and the EPIC detectors (based on the general cross-calibration agreement seen between the different EPIC detectors themselves \citep[e.g.,][]{2014Read}. In these fits we also include a \diskbb\ component (which represents a standard thin accretion disk) in order to model the UV continuum, and assume that the temperature of the \diskbb\ component is the seed photon temperature for the various comptonization components included in our models. Unfortunately, owing to the fact that we only have data for two OM filters which largely overlap in wavelength, the addition of the OM data does not change the best-fit statistics significantly, and the \diskbb\ parameters are poorly constrained. We therefore conclude that the OM data do not change the overall picture implied by the X-ray data in this case, and focus our discussion on the results from the X-ray data alone.

\section{Discussion}\label{sec:discussion}
In the previous sections we presented the changing properties of the AGN \eso, discussed the simple spectral fits of three remarked epochs during \xmm\ observations, and found the variability mainly comes from the entire broadband spectrum, although there is a hint that the soft excess is less variable than the power-law continuum.

We have carried out spectral analyses based on two different hypotheses to explain the soft excess: the warm corona and the relativistic reflection scenario (and their combination). In the process of fitting with the warm corona model, the introduction of the \borus\ model for the distant reflection yields a significant improvement in the model fitting to the data. The \borus\ component models the distant reflection materials in a 3-dimensional geometry, implying that geometrical considerations are important to reproduce the X-ray spectrum of this source. On the other hand, assuming a lamppost geometry for the inner disk reflection enables us to measure the distance between the hot corona and the black hole. It indicates that the hot corona is close to the black hole ($h<5\,R_\mathrm{Horizon}$). Both models provide somewhat satisfactory fits to the spectra. Moreover, the hybrid model {\it Model~C} compensates the lack of the disk reflection in the pure warm corona interpretation, presenting the best statistics and intermediate physical parameters.

Note that the aim of this paper is to compare the warm corona and the relativistic reflection explanations for the soft excess, rather than give a final answer. We thus concentrate on the the physical aspects of {\it Model~A} and {\it Model~B} fits in the following discussions, and put the hybrid model {\it Model~C} as a reference in our discussion.

\subsection{Physical Properties of the Warm Corona Model}\label{sec:discussWC}
In the case of the warm corona, the soft excess originates from the Comptonization of the UV photons, which occurs in a warm ($kT_{e}<1$\,keV) and optically thick ($\tau\sim10\mbox{--}40$) corona above the accretion disk \citep[e.g.,][]{1993Walter,2012Done}. It was estimated that the transition radius between the inner hot corona region and the outer warm corona is at $\sim10\mbox{--}20\,R_\mathrm{g}$ \citep{2013Petrucci,2019Ursini}. In this scenario, most of the gravitational accretion power is proposed to be released into the warm corona rather than in the mid-plane of the accretion disk \citep{petrucci2020}. In addition, some general-relativistic magnetohydrodynamic (GRMHD) simulations indicate that the energy could be transported vertically through the atmosphere within the magnetically dominated accretion disk \citep[e.g.,][]{2009Beckwith,2015Begelman,2019Jiangc}. The warm corona explanation is supported by the observed correlation between the optical/UV and the soft X-ray emission \citep{2011Mehdipour,2018Petrucci}. In the fit of {\it Model~A}, the warm corona model with a hot corona and a torus component describes the observational data well.

However, one noteworthy feature of the warm corona scenario is the relatively hard slope of the continuum $\Gamma^\mathrm{HC}=1.61$, compared with {\it Model~B} and {\it Model~C}. It is much harder than the typical unobscured AGN spectrum $\langle \Gamma \rangle \sim 1.8$ \citep[e.g.,][]{2005Piconcelli,2017Ricci}, and also lower than the previous results obtained by AG14 ($\Gamma\sim1.7\mbox{--}2.1$) through 2005-2010 observations for \eso. But it is worth pointing out that AG14 results were derived by employing a relativistic reflection model to explain the soft excess. \citet{2013Brightman} suggested a correlation between $\Gamma$ and the Eddington ratio $\dot{m}$ by investigating a sample of 69 AGNs. For the Eddington ratio of \eso, we estimate it following the procedure presented by AG14 Section 7.2. We use the total unasborbed $2\mbox{--}10$\,keV luminosity $L_{2-10}\sim3.39\times10^{42}\,\mathrm{erg}\,\mathrm{s}^{-1}$ and assume the X-ray bolometric correction $k_{2-10}=25$ \citep{2009Vasudevan}, which gives $L_\mathrm{Bol}=8.48\times10^{43}\,\mathrm{erg}\,\mathrm{s}^{-1}$. The derived Eddington ratio of $\dot{m}\sim0.015$ is slightly lower than the averaged value $\dot{m}\sim0.02$ reported in AG14. The expected slope of the continuum according to the $\Gamma-\dot{m}$ relation is $\Gamma=1.69\pm0.11$, which is consistent with the results from both scenarios within the uncertainty and the best value is closest to the slope of the hybrid model. \citet{2013Brightman} also caution that such considerable dispersion in the correlation does not make it viable for single sources. Therefore, this estimation only proves that the spectral indices from two scenarios are both possible.

As mentioned before, the warm corona scenario requires a harder continuum in absence of the compensation of the high-energy photons from the inner disk reflection. In our fits, the difference of the photon index between warm corona and relativistic reflection is $\Delta\Gamma\sim0.1$, which can also be found in \citet{2018Garc} and \citet{2019Ursini}. Under the warm corona circumstance, the hot corona produces more high-energy photons than those of the reflection, which means a lower contribution of the reflection component to the high energy regime of X-ray spectrum. Therefore, for the given X-ray AGN spectrum data with the soft excess, the lack of any disc reflection component in the pure warm corona model is likely to provide a harder continuum than the relativistic reflection explanation.

Another irregular feature of the warm corona model is the temperature of the hot corona $kT_\mathrm{e}=20^{+5}_{-3}$\,keV, lower than the values required in {\it Model~B} ($kT_\mathrm{e}>51$\,keV) and {\it Model~C} ($kT_\mathrm{e}=35^{+18}_{-11}$\,keV). Though it is uncommon for most of studied AGNs \citep[e.g.,][]{2015Fabian,2017Fabian,2016Marinucci,2020Balokovi}, \nustar\ has reported several low temperature coronae in recent years, e.g., $\sim25$\,keV \citep[MCG-05-23-016,][]{2015Balokovi}, $\sim50$\,keV \citep{2015Ursini}, $\sim15$\,keV \citep[Ark 564,][]{2017Kara}, $\sim12$\,keV \citep[GRS 1734-292,][]{2017Tortosa}, $\sim40$\,keV \citep[IRAS 00521-7054,][]{2019Walton}. \citet{2018Ricci} also reported a large number of sources with temperature below 50\,keV and found a statistical dependence of the cutoff energy on the Eddington ratio. This correlation suggests that objects with low Eddington ratio ($\dot{m}<0.1$) tend to favor higher cutoff energies, with a mean value of $E_\mathrm{cut}\sim370$\,keV, based on the \swift/BAT 70-months survey of a sample of 838 AGNs. The energy cutoff was usually estimated by the temperature of the hot corona with a relation, $E_\mathrm{cut}\sim(2\mbox{--}3)\,kT_\mathrm{e}$ \citep{2001Petrucci}. Although \citet{2019Middei} reported that this relationship only works for low temperatures, this correction should therefore be appropriate for the warm corona model. Given that the Eddington ratio of \eso\ is not very large ($\dot{m}\sim0.015$), the low electron temperature derived with {\it Model~A} is in conflict with the trend found by \citet{2018Ricci}, while our result of a higher coronal temperature in the relativistic reflection scenario ({\it Model~B}) is compatible with their analysis.

In our fit, the corresponding optical depth of the warm corona ($\tau=34^{+3}_{-2}$), calculated based on the formula~(1) of \citet{1987Lightman}, is within the prediction of \citet{2019Ursini} ($\tau\sim10\mbox{--}40$), while the hot corona ($\tau\sim5$) is optically thick in the warm corona prescription, which is contradictory to the presumed optically-thin hot plasma responsible for the X-ray primary continuum. \citet{2019Garc} examined the effect of photoelectric absorption in scattering media by considering the coronal and photoionization equilibrium, and found that it is dominant in optically-thick regions, which is not taken into account by the warm corona models. They predicted that a forest of absorption lines should be observed in the spectra if the soft excess originates from the warm corona, as the opacity should be dominated by atomic absorption.

Nonetheless, \citet{petrucci2020} and \citet{2020Ballantyne} carried out new simulations and proposed that as long as there is an energy source providing sufficient heating for the warm corona, the absorption features would not be seen, and featureless spectra similar to the soft excess would instead be observed. In these cases, the opacity of the warm corona is dominated by electron scattering, rather than absorption, hence the lack of atomic features. This means that detectable line emission should probably also not be seen from this region, hence our treatment of the warm corona radius as the innermost boundary for the reflection in the hybrid model ({\it Model~C}). Under these conditions, it is unlikely that the warm corona could produce line emission itself, and even if line emission is produced by a standard accretion disc below the warm corona, the optical depth of this region would just scatter the line emission back into the continuum as it emerged \citep{2001Petrucci}.

However, through recent radiative transfer computation in hydrostatic and radiative equilibrium, \citet{2020Gronkiewicz} has shown that for lower accretion rates, thermal instability should prevent the warm corona from forming. The low accretion rate system is unable to provide enough energy to sustain a warm corona \citep{2020Ballantyneb}. It is therefore unclear whether a strong warm corona can be sustained at the low accretion rates relevant here ($\dot{m}\sim0.015$). This may imply that even if a warm corona is present, a contribution from disc reflection would be necessary to produce a strong observed soft excess. 

Note that in {\it Model~C}, the inclusion of the relativistic reflection in the warm corona explanation is able to obtain a hotter and optically thinner hot corona. {\it Model~C} provides a moderate statistical improvement over both {\it Model~A} and {\it B}, and in the case of {\it Model~A}, moves the hot corona parameters into a more physical regime. However, this does not answer the physical questions highlighted previously on whether a warm corona can reasonably exist in \eso. Although the estimation of the flux implies that the soft excess mainly comes from the warm corona in this model, the pure relativistic reflection model is also able to obtain a more physical hot corona without the  warm corona component. Hence, while the warm corona scenario requires the inclusion of a disc reflection component in order to get reasonable parameters for the hot corona, the relativistic reflection model does not require a warm corona component to obtain physically consistent parameters.

\subsection{Physical Properties of the Relativistic Reflection}\label{sec:discussRR}
If the X-ray reflection occurs in close proximity to the black hole, the fluorescent features will be gravitationally blurred to be a smoothed broad spectrum below 2\,keV. \citet{2016Garc} demonstrated that if the disk density is higher than the typically fixed value $n_\mathrm{e}=10^{15}\,\mathrm{cm}^{-3}$, the main effect is the enhancement of the reflected continuum at low energies, further enhancing the soft excess. Below we discuss the relativistic reflection interpretation for the soft excess.
%
\begin{figure}[ht!]
\centering
\includegraphics[width=1\textwidth,width=0.49\textwidth,trim={0 0.0cm 0 0}]{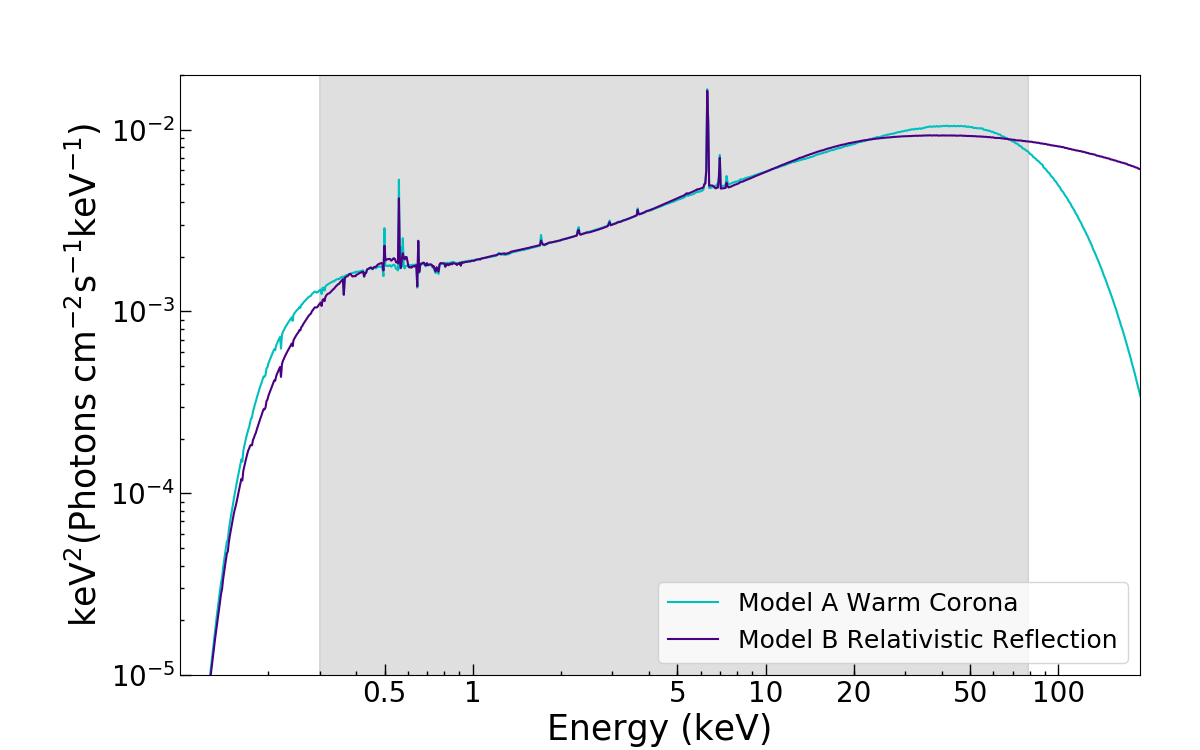}
\caption{
The comparison of the  warm corona ({\it Model~A}) and the relativistic reflection ({\it Model~B}) explanation. The {\it grey} region is the analyzed energy band ($0.3\mbox{--}79.0$\,keV). 
}
\label{fig:comparison}
\end{figure}

We note that the relativistic model has a similar statistical result to the warm corona model for this source. To investigate the extent of the similarity, we compare the total model of {\it Model~A} and {\it Model~B} in Figure~\ref{fig:comparison}. The analyzed energy band is marked as the shaded region. Over this bandpass, the two models can be seen to be very similar, except for the small deviation above 20\,keV. The deviation and the unconstrained temperature of the hot corona ($kT_\mathrm{e}>51\,$keV) may be the results of the limitation of \relxilllpD\ ($E_\mathrm{cut}=300\,$keV). However, this is difficult to test since currently there is no official version of \relxilllpD\ with a variable high energy cut-off. Outside the coverage, differences can be seen at softer energy bands. The warm corona interpretation requires slightly more photons in the soft bands ($<0.5$\,keV) compared with the relativistic reflection model. Consequently, improved theoretical models and future observations covering wider energy bands with high-resolution are essential for exploring the nature of the soft excess.

The relativistic reflection model requires a compact corona ($h=3.0^{+1.0}_{-0.3}\,R_\mathrm{Horizon}$) and a highly spinning black hole ($a_\star>0.927$). While the real configuration of the hot corona is usually unknown, assuming a lamp post geometry, where the source is a point in the rotational axis, is an idealized but effective way to explore the location of the hot corona. Moreover, the iron abundance is close to the solar value $\mathrm{A}_\mathrm{Fe}=1.5^{+0.4}_{-0.3}$. Several recent analyses of both active galaxies and X-ray binaries have found that allowing for higher densities for the disk, as predicted by standard accretion theory \citep{1994Svensson} in many cases, can be important with regards to the iron abundance inferred and in particular avoiding highly super-solar abundances \citep[e.g.,][]{2018Jiang,2018Tomsick,2018Garc}.

Compared with {\it Model~C}, the distinct differences are the inclination angle ($i\sim32^\circ$) and the ionization parameter (\logxi$=0.4^{+0.3}_{-0.1}$). Since \eso\ is an unobscured AGN, the inclination angle is expected to be low to prevent the obscuration through the torus. Both the face-on and low inclination angle solutions therefore cannot be ruled out. The low ionization parameter of the relativistic reflection component indicates that the degree of ionization on the accretion disk is relatively low. To confirm this result, we estimate the ionization state through its definition $\xi = L_{X}/n_{e}R^2$. Assuming $n_{e}=10^{18.3}\,\mathrm{cm}^{-3}$, $R=3\,R_{g}$, and the X-ray luminosity measured at $L_{X}=1.52\times10^{43}\,\mathrm{erg}\,\mathrm{s}^{-1}$, we derive $\xi\sim0.02$\,erg\,cm\,s$^{-1}$, implying a low ionization state. The previously reported ionization states of \eso\ required from just a relativistic reflection model show a low ionization state and at most up to $\xi\sim35$\,erg\,cm\,s$^{-1}$, consistent with our fitting result. According to \citet{2011Ballantyne}, they found a positive statistical correlation between $\xi$ and the AGN Eddington ratio $\dot{m}$ based on the simple $\alpha$-disk theory. Thus, AGNs with low Eddington ratio are expected to exhibit less ionized inner accretion disks. The physical interpretation is that the accretion rate affects the fraction of the accretion energy dissipated in the corona \citep[e.g.,][]{1994Svensson,2002Merloni,2009Blackman}, which emits X-ray photons to photoionize the inner disk surface. The estimation of the ionization parameter through plugging the Eddington ratio $\dot{m}=0.015$ into Formula~(1) of \citet{2011Ballantyne}, is \logxi$\sim1.3$, larger than {\it Model~B} result but identical with {\it Model~C}. The reason for the low ionization state in {\it Model~B} is possibly due to the high gas density. \citet{2016Garc} reported that the high densities would enhance the collisional de-excitation and when the density is high enough (at \logne$\sim19$), the three-body recombination becomes important, which lowers the ionization state of the disk.

The relativistic reflection explanation requires a dense accretion disk (\logne$>18.3$). This result is consistent with previous findings that larger gas density than the previously adopted value of $\log{[n_\mathrm{e}/\mathrm{cm}^{-3}]}=15$, is usually required for SMBHs with $\log{[m_\mathrm{BH}/M_\odot]}\lesssim8$, like \eso\ \citep[$\sim7.7$;][]{2019Jiang}. Another factor that affects the expected disk density is the accretion rate. According to the $\alpha$-accretion disk \citet{1973Shakura} model, \citet{1994Svensson} derived a relationship between the density of a radiation-pressure-dominated disk and the accretion rate, $\log{[n_\mathrm{e}]}\propto-2\log{[\dot{m}]}$, suggesting the lower accretion rate leads to a higher gas density. The specific formula of the relationship \citep[i.e., Equation 8 in][]{1994Svensson} is,
\begin{equation}
\small
    n_\mathrm{e} =\frac{1}{\sigma_\mathrm{T}R_\mathrm{S}}\frac{256\sqrt{2}}{27}\alpha^{-1}r^{3/2} \dot{m}^{-2}\left[ 1-(R_\mathrm{in}/r)^{1/2}\right]^{-2} \left[\xi(1-f)\right]^{-3}
\end{equation}
where $\sigma_\mathrm{T}=6.64\times10^{25}\,\mathrm{cm}^2$ is the Thomson cross section; $R_\mathrm{S}\equiv2GM_\mathrm{BH}/c^2$ is the Schwarzchild radius; $\alpha\approx0.1$ is the standard \citet{1973Shakura} viscosity parameter; $r$ is in the units of $R_\mathrm{S}$; $R_\mathrm{in}$ is the inner radius of the black hole, usually assumed at the ISCO; $\xi$ is chosen to be unity in the radiative diffusion equation; and $f$ is the fraction of the total transported accretion power released from disk to the hot corona. 
%
\begin{figure}[ht!]
\centering
\includegraphics[width=1\textwidth,width=0.49\textwidth,trim={0 0.0cm 0 0}]{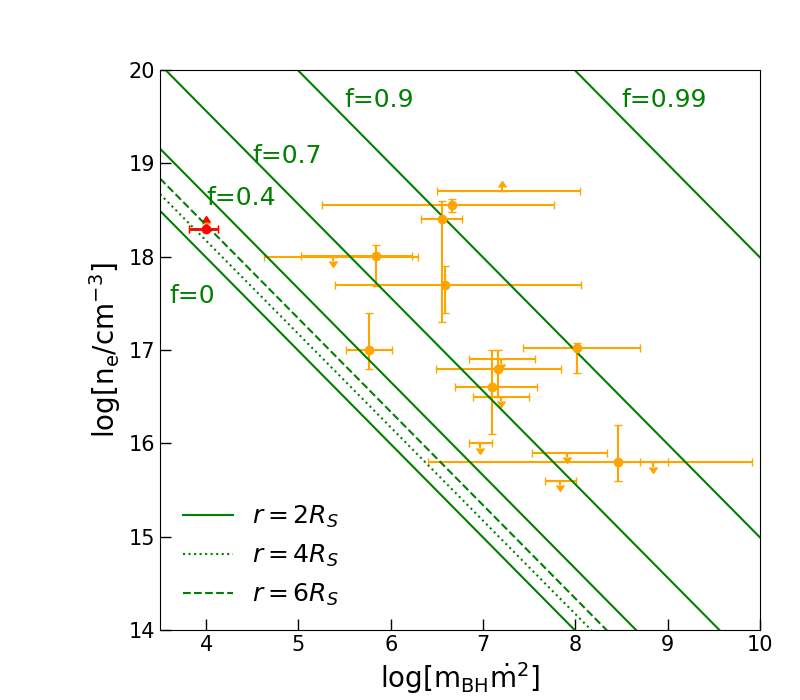}
\caption{Disk gas density $\log{[n_\mathrm{e}]}$ versus $\log{[m_\mathrm{BH}\dot{m}^{2}]}$. The {\it orange} points are the AGN samples considered in \citep{2019Jiang} and the {\it red} point is \eso. The solid {\it green} lines are the solutions for disk density at $r=2\, R_{S}$ for different values of $f$ \citep{1994Svensson}, assuming the inner radius at ISCO $R_\mathrm{in}=1.24\,R_\mathrm{g}$ of a maximally rotating Kerr BH ($a_\star=0.998$). The dotted and dashed lines are for $r=4,6\,R_{S}$ and $f=0$.
}
\label{fig:mmdotne}
\end{figure}

To compare with other sources fitted with the high-density reflection model and place \eso\ in the context of the \citet{1994Svensson} model, we reproduce Figure~4 in \citet{2019Jiang}, plotted in Figure~\ref{fig:mmdotne}, which is a diagram of the disk density $n_\mathrm{e}$ versus $\log{[m_\mathrm{BH}\dot{m}^{2}]}$ based on a sample of Seyfert 1 galaxies analyzed in their paper, marked as orange points, and pointed out the location of \eso\ with red. The difference is that they assume a black hole with $a_\star=0.95$ while we assume a maximally spinning black hole $a_\star=0.998$ and thus the $R_\mathrm{in}=1.24\,R_\mathrm{g}$. The solid green lines show the analytic solution of disk density at $r=2\,R_\mathrm{S}$ for different $f$. The dotted and dashed lines correspond to the disk density at $r=4,6\,R_{S}$ for $f=0$. \eso\ is near the edge of the solution at $r=2\,R_\mathrm{S}$ for $f=0.4$, compatible with the correlation predicted by \citet{1994Svensson} that high disk density tends to occur in the AGN with low $\log{[m_\mathrm{BH}\dot{m}^{2}]}$. The solution crossing the optimal value of \eso\ suggests, in classical disk theory, if $\sim25\%$ of the accretion power in the disk is released to the corona, $r<2\,R_\mathrm{S}$ region of the disk is dominated by radiation pressure, or another possibility is that if the radiation-pressure-dominated area is larger (e.g., $r<6\,R_\mathrm{S}$), then lower $f$ could be expected. It should be noticed that the disk density in the reflection model is the surface density of the disk while the density parameter in the $\alpha$-disk model is really the density at the mid-plane, but here we assume the disk is vertically uniform, since the structure of the accretion disk is not well understood. Moreover, the GRMHD simulations from \citet{2019Jiangc}, which are the closest to \eso\ at $\dot{m}=7\%\mbox{--}20\%$, show that the photospheric and mid-plane densities only differ by a factor of a few ($<10$) within $\sim10\,R_\mathrm{g}$ of the black hole.

\subsection{Simulations of the future missions}
%
\begin{figure}[ht!]
\centering
\includegraphics[width=1\textwidth,width=0.49\textwidth,trim={0 0.0cm 0 0}]{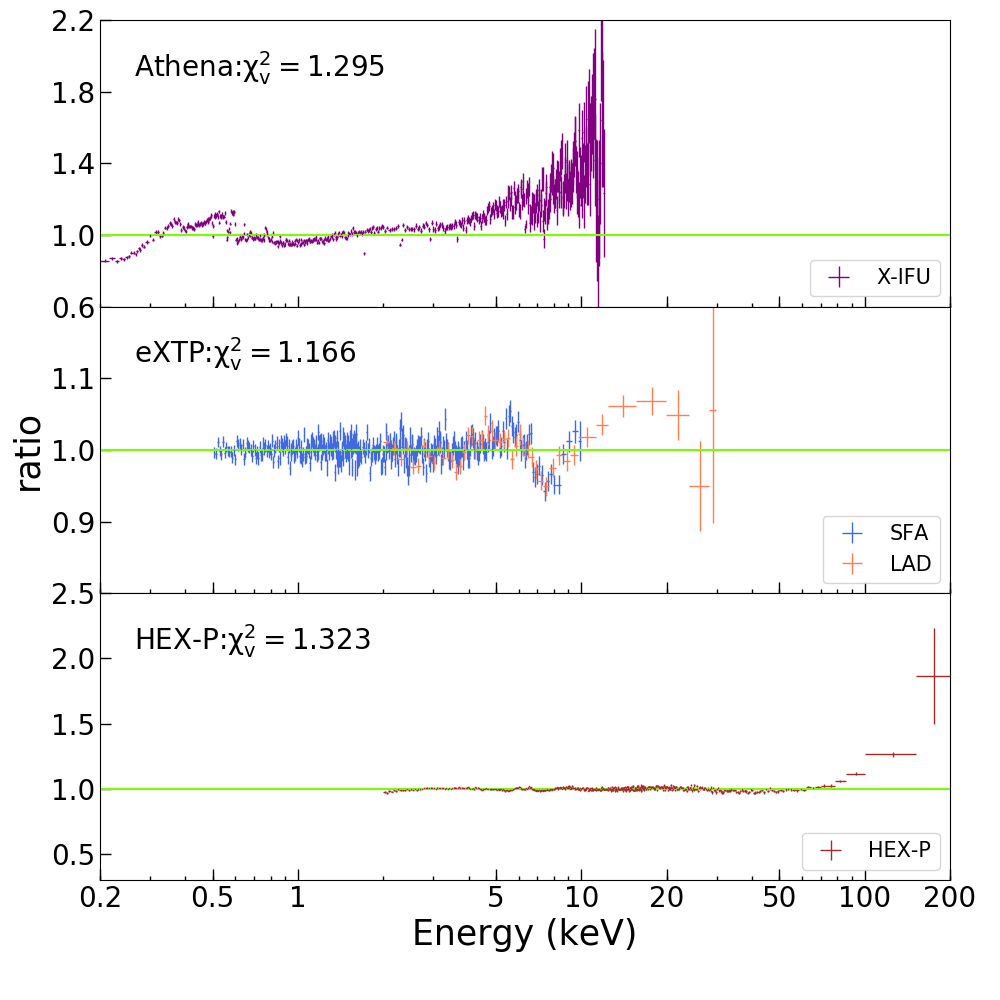}
\caption{
{\it Upper to Bottom: }Data-to-model ratios of the {\it Model~A} fits with the \athena, \eXTP\ and \hexp\ data simulated by using {\it Model~B}. 
}
\label{fig:simulation}
\end{figure}

To estimate the extent to which future missions can distinguish between the two main models proposed to explain the soft excess (i.e., warm corona or relativistic reflection), we utilize the \texttt{fakeit} task in {\sc xspec} to generate simulated data of future missions by using {\it Model~B} with the exact best-fit values, and use {\it Model~A} to fit the fake data. We choose the {\it Advanced Telescope for High-Energy Astrophysics} \citep[{\it Athena},][]{2013Nandra}, the {\it enhanced X-ray Timing and Polarimetry mission} \citep[{\it eXTP},][]{2016Zhang} and the {\it High-Energy X-ray Probe} \citep[{\it HEX-P},][]{2018Madsen}. The X-ray Integral Field Unit \citep[X-IFU,][]{2016Didier} instrument on \athena\ has an effective energy range of $0.2\mbox{--}12$\,keV with $2.5$\,eV spectral resolution, satisfying our requirement to make a distinction. The Spectroscopic Focusing Array (SFA) and the Large Area Detector \citep[LAD,][]{2018Feroci} on \eXTP\ offers a better spectral resolution than \xmm\ and \nustar\ ($180$\,eV and $250$\,eV, respectively). The concept probe mission  \hexp\ provides a broadband ($2\mbox{--}200$\,keV) response and a much higher sensitivity than any previous mission in the hard energy band. We assume an exposure time of $100$\,ks for both the source and background spectra. The data-to-model ratios and the fitting statistics are plotted in Figure~\ref{fig:simulation}. We note that the deviations above 100\,keV in the case of \hexp\ are too large to be presented. All of the fits show clear discrepancies between the warm corona and the relativistic reflection models, where the main difference comes from the lower and higher energies and the iron complex region. This implies that the nature of the soft excess is likely to be determined by the high-quality observations performed by future telescopes.

\section{Conclusions}\label{sec:conclusion}
We have performed spectral fits to \xmm\ ($\sim121\,\mathrm{ks}$) and \nustar\ ($\sim102\,\mathrm{ks}$) simultaneous observations of the Seyfert 1.5 galaxy \eso\, observed in 2016. The spectrum presents an obvious soft excess, the X-ray reprocessing features from an illuminated accretion disk, and a primary non-thermal X-ray continuum. We have studied the origin of the soft excess by fitting either with a warm and optically thick corona, with a relativistically blurred high-density reflection model, or their combination. By investigating the variability during the observations, we find that any spectral variability is subtle at best, although there is a hint that the soft excess is less variable than the power-law continuum.

Although the warm corona and relativistic reflection explanations cannot be uniquely distinguished from statistics, the model fits presented here and the physical conditions relevant for \eso\ (i.e. its moderately low accretion rate) may argue in favour of the reflection-dominated scenario. In the context of the warm corona, a low-temperature ($kT_\mathrm{e}\sim20\,\mathrm{keV}$) and optically thick ($\tau\sim5$) hot corona are uncommon for a typical unobscured AGN. Although in the hybrid model, where the hot corona is hotter and optically thinner and the soft excess flux is dominated by the warm corona, this model is at the cost of complexity and a simpler pure relativistic reflection model can also obtain a similar result. In addition, the energy source for the warm corona requires a sufficient local dissipation of accretion power and a magnetically-dominated disk, but this configuration may not be stable for the low accretion rate ($\dot{m}\sim0.015$) of this system. On the other hand, the physical implications of the relativistic reflection explanation, such as the spectral index ($\Gamma\sim1.75$) and the corona temperature ($kT_\mathrm{e}>51\,\mathrm{keV}$), are more reasonable. For \eso, it also requires a compact corona ($h\sim3\,R_\mathrm{Horizon}$), a highly spinning ($a_\star>0.927$) black hole and a high-density (\logne$\gtrsim18.3$) accretion disk with low ionization state (\logxi$\sim 0.4$). Therefore, given the discussions presented before, we tentatively favor the high-density relativistic reflection interpretation as an explanation for the soft excess.

These two scenarios can be further distinguished by future work on two fronts. One is to obtain broader energy band and higher-quality simultaneous observations of Seyfert galaxies, since the difference between two models are significantly much more evident outside the energy coverage of current observatories. Future missions such as {\it Athena} \citep{2013Nandra}, {\it XRISM} \citep{2018Tashiro}, and {\it Lynx} \citep{2018Ozel}, are crucial for providing reliable high-resolution X-ray observations. Our simulations of the spectral data by using \athena, \eXTP, and \hexp\ response files exhibit evident discrepancies between the two models. The other aspect concerns studies of the timing properties, such as X-ray reverberation lags \citep[e.g.,][]{2009Fabian,2013Marco,2016Kara}. It is an independent method to study the short-term variability of the disk and could test the presence of the time lag between the continuum and the soft excess emission, which is the signature predicted by the relativistic reflection scenario \citep[e.g.;][]{2019Ingram, 2019Mastroserio, 2020Vicentelli}. The future enhanced X-ray Timing mission {\it eXTP} \citep{2016Zhang} will be particularly suited for such studies, which are expected to confirm the relativistic reflection explanation of the soft excess in AGN.

\section{Acknowledgements}
We thank to the referee for their constructive comments. J.A.G. acknowledges support from NASA grant NNX17AJ65G and from the Alexander von Humboldt Foundation. He is also a member of Teams 458 and 486 at the International Space Science Institute (ISSI), Bern, Switzerland, and acknowledges support from ISSI during the meetings in Bern. D.J.W. acknowledges support from STFC through an Ernest Rutherford fellowship. R.M.T.C. has been supported by NASA grant 80NSSC177K0515. This work was partially supported under NASA contract No. NNG08FD60C and made use of data from the NuSTAR mission, a project led by the California Institute of Technology, managed by the Jet Propulsion Laboratory, and funded by the National Aeronautics and Space Administration. We thank the NuSTAR Operations, Software, and Calibration teams for support with the execution and analysis of these observations. This research has made use of the NuSTAR Data Analysis Software (NuSTARDAS), jointly developed by the ASI Science Data Center (ASDC, Italy) and the California Institute of Technology (USA).

{\it Facilities}: \nustar\ \citep{2013Harrison}, \xmm\ \citep{2001Jansen}, \athena\ \citep{2013Nandra}, \eXTP\ \citep{2016Zhang}, \hexp\ \citep{2018Madsen}.

{\it Software}: \xillver\ \citep{2010Garc,2013Garc}, \relxillD/\relxilllpD\ \citep{2014Dauser,2014Garc,2016Garc}, \borus\ \citep{2019Balokovi}.

\bibliographystyle{aasjournal}
\bibliography{ref}

\end{document}